\documentstyle[11pt,psfig]{article}

\newcommand{\mysection}{\setcounter{equation}{0}\section}

\begin{document}

\title{\vskip -1.3in
\hfill {\small ITP-SB-95-08}
\vskip  0.6in
{\Large \bf Heavy-quark correlations in deep-inelastic electroproduction}}
\author{{\bf B. W. Harris and J. Smith} \\ \\ {\small \em Institute for
Theoretical Physics}, \\ {\small \em State University of New York at
Stony Brook}, \\ {\small \em Stony Brook, New York 11794-3840, USA}}
\vskip  1.0cm
\date{March 1995}
\maketitle
\centerline{Submitted to Nuclear Physics B}
\vskip 1.0cm

\begin{abstract}

We have completed the next-to-leading order perturbative
QCD corrections to the
virtual-photon exclusive differential cross sections for heavy quark
production in deep-inelastic electron-proton scattering, i.e.
$e + P \rightarrow Q + \overline{Q} + X$.  Using these results, we
have computed distributions which are sensitive to correlations among
the heavy quark, the heavy antiquark, and the associated jet.
Some predictions for charm and bottom heavy quark
production at the electron-proton collider HERA are presented.

\end{abstract}

\newpage

\mysection{Introduction}
Order $\alpha_s$ QCD corrections to structure functions containing
heavy quarks and to single heavy quark inclusive distributions
in deep-inelastic electroproduction
( ${\em i.\ \! e.\ \! }$, $\gamma^{\ast}(q) + P(p) \rightarrow Q(p_1)
+ X$ where $X$ stands for any final hadronic
state allowed by quantum-number conservation and $P(p)$ is a
proton of momentum $p$ )
were recently published in \cite{LRSvN1} and \cite{LRSvN2},
respectively.  By combining the next-to-leading order (NLO)
heavy quark structure functions with the corresponding
light-quark structure functions \cite{ZvN}, the heavy quark
content of the nucleon has been studied as a function of
$Q^2 = -q^2$ and $x=Q^2/2p \cdot q$ \cite{LRSvN3}.
Event rates for charm production integrated over bins
in $x$ and $Q^2$ relevant to HERA data have been calculated in
\cite{RSvN}.

To further the study of deep-inelastic electroproduction of
heavy quarks we have recalculated the virtual-photon-parton cross
sections of \cite{LRSvN1} in an exclusive fashion.
This enables us to study the single and double differential
distributions and correlations among all outgoing particles in the
virtual-photon induced reaction
$\gamma^{\ast} + P \rightarrow Q + \overline{Q} + X$ with $X=0$
or $1$ jet and to easily incorporate experimental cuts.
A short letter, containing a study of invariant mass distributions of
heavy-quark-heavy-antiquark pairs, has been published \cite{HS}.
In this paper we present all the calculational
details and additional distributions.
The transverse and longitudinal photon components are
treated separately and the
latest CTEQ3 parton densities \cite{CTEQ}, consistent with the
newly released HERA data \cite{HERA},
are used in the kinematic regime appropriate for production of charm and
bottom quarks at HERA. We make our predictions
at fixed values of $Q^2$ $(\geq 8.5 ({\rm GeV/c})^2)$
and $x$ $(\geq 4.2 \times 10^{-4})$.

Our treatment of electroproduction at HERA extends the
existing studies of inclusive QCD corrections in the virtual-photon
channel \cite{LRSvN1}, inclusive QCD corrections in the real-photon
channel \cite{SvN}, and exclusive QCD corrections
in the real-photon channel \cite{frix}, allowing for an extensive
comparison with present and future experimental data.  Heavy quark
electroproduction has already played an important role in the
determination of the gluon distribution function in the proton at low $x$.
In fact, a study of the photoproduction and the Weizs\"{a}ker-Williams
electroproduction of charmed mesons has just appeared \cite{dstar}.
Production cross sections of charm and bottom quarks are also relevant
in the proposed study of the Cabibbo-Kobayashi-Maskawa (CKM)
matrix elements through the rare decays of $D-$ and $B-$
mesons and the $D\overline{D}$ and $B\overline{B}$ mixing \cite{Sch}.
Our study of deep-inelastic electroproduction is motivated by our
desire to avoid complications associated with the parton densities
of the photon which must be incorporated when $Q^2$ is small.
Therefore our results should give a cleaner
test of NLO perturbative QCD\@.  By examining distributions and correlations
that are trivial at lowest order (for example, the azimuthal angle between
the heavy quark and the heavy antiquark) one directly tests the hard
scattering cross section that is predicted by perturbative QCD and
factorization.  Here we should note that this is a fixed order
perturbative calculation and suffers from the same problems as all NLO
calculations.  Hence there are regions in phase space
where it will break down.
For example, in the above mentioned azimuthal angle distributions,
if one looks at the prediction too near the back-to-back
configuration there will be an extra enhancement of the cross section due to
multiple soft gluon emission which is not included in our NLO result.

We stress that here we only consider {\em extrinsic} heavy quark
production, involving Bethe-Heitler and Compton
production from massless partons.  For a discussion
of {\em intrinsic} production, where the heavy quark is considered to
be part of the proton's wavefunction, see Brodsky {\em et al}.\ \cite{int}.
A variable flavor scheme which joins the extrinsic
heavy flavor production picture at $\mu_{\rm phys} \ll m$ with a light mass
parton density picture at $\mu_{\rm phys} \gg m$ has
been discussed by Aivazis {\em et al}.\ \cite{var}.
By comparing the fixed flavor scheme calculation
of \cite{LRSvN1} with the variable flavor scheme
of \cite{var} it is concluded in \cite{comp} that the
former yields the most stable and reliable results for
$F_2(x,Q^2,m^2)$
in the threshold region ( ${\em i.\ \! e.\ \! }, \, Q^2 \leq 10m^2$
where $m$ is the mass of the heavy quark).

In our exclusive computation we use the subtraction method which is
based on the replacement of divergent (collinear or soft)
terms in the squared matrix elements by generalized plus distributions.
This allows us to isolate the soft and collinear
poles within the framework of dimensional regularization without
calculating all the phase space integrals
in a spacetime dimension $n\ne 4$ as usually required in
a traditional single particle inclusive computation.
The method has been used in the context
of electron-positron annihilation \cite{ellis},
hadroproduction of $Z^0$ boson pairs \cite{mele},
hadroproduction of jets \cite{ks},
hadroproduction of heavy quarks \cite{mnr},
photoproduction of heavy quarks \cite{frix}, and
hadroproduction of a $W$ boson plus a photon \cite{ms}.
The essence of the method is nicely
described and compared to the also popular phase-space slicing method
\cite{owens} in the introduction of the paper by Kunszt and Soper \cite{ks}.
The paper of Mangano {\em et al}.\ \cite{mnr} contains many useful details.
In this method the expressions for the squared matrix elements in the
collinear limit appear in a factorized form, where poles in $n-4$ multiply
splitting functions and lower order squared matrix elements.
The cancellation of collinear singularities is then performed using the
factorization theorem \cite{CSS}.
The expressions for the squared matrix elements in the soft limit appear
in a factorized form where poles in $n-4$ multiply lower
order squared matrix elements.  The cancellation of soft singularities
takes place upon adding the contributions from the renormalized
virtual diagrams.
Since the final result is in four-dimensional space time, we can compute all
relevant phase space integrations using standard Monte Carlo
integration techniques \cite{lepage} and produce histograms
for exclusive, semi-inclusive, or inclusive  quantities
related to any of the outgoing particles.  We can also
incorporate any reasonable
set of experimental cuts.  Our computer code has no small phase space slice
parameters and the parameters defining the generalized plus distributions
may be tuned to give fast numerical convergence,
which is an advantage of using this subtraction method.

We therefore have a new, and more general,
calculation of the scale independent (Wilson) coefficient functions
(or scale independent partonic cross sections)
$c^{(l)}_{k,i}(\eta,\xi)\, , \, \bar c^{(1)}_{k,i}(\eta,\xi)$, and
$d^{(1)}_{k,i}(\eta,\xi)$ defined  in \cite{LRSvN1},
as functions of $\eta=s/4m^2-1$, where $s$ is the square of
the c.\ m.\ energy in the virtual-photon-parton system, and $\xi=Q^2/m^2$.
We checked the $\eta$ and $\xi$ dependence of
these scale independent coefficient functions and the $x$ and $Q^2$
dependence of the hadronic structure functions
$F_2(x,Q^2,m^2)$ and $F_L(x,Q^2,m^2)$,
obtained after convolution of the coefficient functions with the
light quark and gluon densities in the proton, against the results in
\cite{LRSvN1}. We comment on this comparison below.
The results for the single quark inclusive transverse momentum and rapidity
distributions,  $dF_{k}(x,Q^2,m^2,p_t)/dp_{t}$ and
$dF_{k}(x,Q^2,m^2,y)/dy$ $(k=G,L)$ respectively,
were also checked against the results in \cite{LRSvN2} to
make sure that all our computer programs yield  consistent results.

In sect.\ 2 we introduce our notation used in the calculation of
the cross sections for the off-shell photoproduction of heavy quarks.
We closely follow the notations of \cite{LRSvN1} and \cite{frix}.
In sect.\ 3 we discuss the analytical results needed for the calculation
of the numerical results.  The latter are presented in sect.\ 4.  We
concentrate on those plots which are dependent on the emission
of an extra parton (or jet) and only show single differential distributions.
Our conclusions are presented in sect.\ 5.
Kinematics and other useful results are relegated
to appendix A.
Appendix B contains a discussion of the soft limit of
the matrix elements.
Appendix C contains a discussion of the collinear limit of
the matrix elements.
We have included these appendices to help with the identification of the
various pieces of our computer code.

\mysection{Notation and Born cross section}
In this section we introduce our notation and define the quantities which
we will calculate.  We then review the Born level results.
The kinematics and other useful results are given in
appendix A.  Following \cite{LRSvN1}, we denote the parton subprocesses
contributing to heavy quark production in deep-inelastic electron-hadron
scattering by (see fig.\ 1)
\begin{eqnarray}
\label{reaction}
\gamma^{\ast} (q) + a_1(k_1) \rightarrow Q(p_1) + \overline{Q}(p_2) +
a_2(k_2) + \cdots + a_j(k_j),
\end{eqnarray}
where the photon momentum is space-like ( $q^2 = -Q^2 < 0$ ), $a_i(k_i)$
stand for massless partons ($k_i^2=0$), and $Q \, (\bar{Q})$ is the heavy
(anti) quark ($p_1^2=p_2^2=m^2$).  We consider the partonic structure tensor
\begin{equation}
\label{Wdef}
W_{\mu\nu}=\frac{1}{2 s^{\prime}} \frac{1}{2} \frac{1}{1+
\delta_{g a_1} \epsilon/2}
K_{a_1 \gamma} \left[ \sum M_{\mu}(2) M_{\nu}^{\ast}(2) d \Gamma_2 +
\sum M_{\mu}(3) M_{\nu}^{\ast}(3) d \Gamma_3 + \cdots \right],
\end{equation}
where each term in the square brackets corresponds to a perturbative
expansion in $\alpha_s$.  The terms in the structure tensor are the
flux factor $1/2 s^{\prime}$, the initial degree of freedom average
$1/2(1+ \delta_{g a_1} \epsilon / 2)$ where $\delta_{g a_1}$ is the
Kronecker delta function,  the color average $K_{a_1 \gamma}$,
the 2 to $j$ body amplitude $M(j)$, the $j$ body phase space $d \Gamma_{j}$,
and the sum over initial and final degrees of freedom $\Sigma$.
The color average factor has the specific values
$K_{g \gamma} = 1/(N^2-1)$ and
$K_{q \gamma} = K_{\bar{q} \gamma} = 1/N$, where the number of colors is
$N=3$ for $SU(3)$.  We work in the context of dimensional regularization with
space time dimension $n=4+\epsilon$.  Since we integrate over the azimuthal
angle between the plane containing the incoming and outgoing leptons and the
plane defined by the incoming parton and outgoing heavy quark the partonic
tensor (\ref{Wdef}) only has two terms which can be written as
\begin{eqnarray}
\label{Wdecomp}
W_{\mu\nu} & = & d \sigma_T \left( -g_{\mu \nu} + \frac{q_{\mu}
	q_{\nu}}{q^2} \right) \\ \nonumber
& & +\left(k_{1\mu}-\frac{k_1 \cdot q}{q^2} q_{\mu} \right)
	\left(k_{1\nu}-\frac{k_1 \cdot q}{q^2} q_{\nu} \right) \left(
	\frac{-4 q^2}{s^{\prime^2}} \right) \left( d \sigma_T
	+ d \sigma_L \right).
\end{eqnarray}
Using the projection operators $g^{\mu \nu}$ and $k_1^{\mu} k_1^{\nu}$
the partonic cross sections can be written as follows:
\begin{eqnarray}
\label{sigG}
d \sigma_G & = & -\frac{1}{2} \frac{1}{1+\epsilon/2} g^{\mu \nu}
W_{\mu \nu}, \\
\label{sigL}
d \sigma_L & = & -\frac{4 q^2}{(s^{\prime})^2} k_1^{\mu} k_1^{\nu}
		W_{\mu \nu}.
\end{eqnarray}
We work with $d \sigma_G$ and $d \sigma_L$ for convenience and later use
these to get the transverse partonic cross section $d \sigma_T$ via the
relation
\begin{equation}
\label{sigT}
d \sigma_T = d \sigma_G + \frac{1}{2} \frac{1}{1+\epsilon/2} d \sigma_L.
\end{equation}
For each of the three partonic channels $a_1=g,q,\bar{q}$ we expand the
corresponding parton cross sections in terms of the number of outgoing
particles in reaction (\ref{reaction}):
\begin{equation}
\label{sig}
d \sigma_{i,a_1} = d \sigma_{i,a_1}^{(2)} + d \sigma_{i,a_1}^{(3)} + \cdots
\end{equation}
where $i=G,L$ and
\begin{equation}
\label{sigj}
d \sigma_{i,a_1}^{(j)} = C_{i,a_1} M^{a_1}_{i} \! (j) \, d \Gamma_j .
\end{equation}
Using eq.\ (\ref{Wdef}) we find
the partonic cross sections are related to the squared matrix elements by
\begin{eqnarray}
\label{M}
M^{a_1}_{G}(j) & = & - g^{\mu \nu} \sum M_{\mu} (\gamma^{\ast} a_1
\rightarrow j) M_{\nu}^{\ast}(\gamma^{\ast} a_1 \rightarrow j), \nonumber \\
M^{a_1}_{L}(j) & = & -\frac{4 q^2}{(s^{\prime})^2}
k_1^{\mu} k_1^{\nu} \sum M_{\mu} (\gamma^{\ast} a_1
\rightarrow j) M_{\nu}^{\ast}(\gamma^{\ast} a_1 \rightarrow j),
\end{eqnarray}
with
\begin{equation}
\label{C}
C_{i,a_1} = \left( \frac{1}{2} \frac{1}{1+\epsilon/2} \right)
a_i(\epsilon)\left( \frac{1}{2 s^{\prime}} \frac{1}{2}
\frac{1}{1+ \delta_{ga_1} \epsilon/2} K_{a_1 \gamma} \right),
\end{equation}
where $a_G (\epsilon)=1$ and $a_L (\epsilon)=2(1+\epsilon/2)$.
Note that
\begin{equation}
\label{Crelation}
C_{i,q} = 2 C_F ( 1 + \epsilon/2 ) C_{i,g} \, ,
\end{equation}
with the color factor $C_F = (N^2-1)/2N = 4/3$ for $SU(3)$.
The expressions for $M^{a_1}_{G}(j)$ and $M^{a_1}_{L}(j)$ were evaluated
using the symbolic algebra program FORM \cite{jos}.

We begin our analysis by considering the lowest order process
contributing to reaction (\ref{reaction}) which is shown
in fig.\ 2.  Namely,
\begin{equation}
\label{born}
\gamma^{\ast}(q) + g(k_1) \rightarrow Q(p_1) + \bar{Q}(p_2).
\end{equation}
The projections of the matrix element for this reaction are
calculated according to eq.\ (\ref{M}).  We calculate in
$n=4+\epsilon$ space-time dimensions. Both the $O(\epsilon)$ and
$O(\epsilon^2)$ terms are needed for a correct treatment
of renormalization and mass factorization.  Summing over
initial and final spins
\begin{equation}
\label{Mborn}
M^{g}_{k}(2) = 8 g^2 e^2 e_H^2 N C_F B_{k,QED},
\end{equation}
with $k=G,L$ and
\begin{eqnarray}
B_{G,QED} & = & \frac{u_1}{t_1}+\frac{t_1}{u_1}+\frac{2q^2s}{t_1 u_1}
+\frac{4 m^2 s^{\prime}}{t_1 u_1} \left(1-\frac{m^2 s^{\prime}}{t_1 u_1}
\right)+\frac{2m^2q^2}{t_1 u_1}\left(2-\frac{ (s^{\prime})^2}{t_1 u_1} \right)
\nonumber \\
& + & \epsilon \left\{ -1 + \frac{(s^{\prime})^2}{t_1 u_1}
+\frac{s^{\prime} q^2}{t_1 u_1}+\frac{q^4}{t_1 u_1}
-\frac{m^2q^2(s^{\prime})^2}{t_1^2 u_1^2}
\right\} + \epsilon^2 \frac{(s^{\prime})^2}{4t_1 u_1}, \\
B_{L,QED} & = & \frac{-4q^2}{(s^{\prime})^2} \left[s-
\frac{m^2 (s^{\prime})^2}{t_1 u_1} \right].
\end{eqnarray}
The kinematic variables are defined in appendix A.
Here $g$ and $e$ are the strong and electromagnetic coupling constants
respectively and $e_H$ is the charge of the heavy quark in units of $e$.
The lowest order cross section now follows directly from eq.\ (\ref{sigj}).

\mysection{Order $\alpha_s$ corrections}
In this section we describe the calculation of the NLO
corrections to the lowest order gluon fusion cross sections which were derived
in the last section.  As mentioned in the introduction, we proceed
via the subtraction method wherein one extracts the soft and/or collinear
singularities without integration over final state particles $Q$ and $a_2$
as done in the previous single particle inclusive calculation
\cite{LRSvN1}, \cite{LRSvN2}.  This allows us to plot correlations
between final state particles which will be presented in sec.\ 4.
We first discuss the gluon-bremsstrahlung reaction in detail and then,
in the following subsection, the light quark contributions.  The
kinematics of the three body final states are discussed in appendix A, which
also contains the definitions of the variables used here.
\subsection{$\gamma^{\ast}g$ Channel}

The gluon-bremsstrahlung reaction, shown in fig.\ 3, is
\begin{equation}
\label{gbrem}
\gamma^{\ast}(q)+g(k_1) \rightarrow g(k_2)+Q(p_1)+\bar{Q}(p_2).
\end{equation}
We begin by writing the partonic cross sections as
\begin{eqnarray}
d \sigma_{i,g}^{(3)} & = & C_{i,g} M_i^g(3) d\Gamma_3 \nonumber \\
		     & = & C_{i,g} f_i^g (x,y,\theta_1,\theta_2)
			\overline{d\Gamma_3},
\end{eqnarray}
where
\begin{equation}
\label{fgi}
f_i^g (x,y,\theta_1,\theta_2) \equiv t^{\prime} (u^{\prime}
	- q^2 s_5/s) M^g_i (3),
\end{equation}
\begin{equation}
\overline{d\Gamma_3} \equiv \frac{d\Gamma_3}{t^{\prime}
\left( u^{\prime}-q^2 s_5/s \right)}.
\end{equation}
We suppress all arguments of $f_i^g$ other than $x,y,\theta_1,\theta_2$
for notational convenience.  These are the variables we integrate over
to calculate the partonic cross sections for reaction (\ref{gbrem}).
The subtraction method proceeds by multiplication of the
squared matrix element, and division of the phase space, by invariants
(or combinations thereof) that vanish in the soft and/or
collinear limit.  The invariants are chosen to cancel the singularities
in the squared matrix elements due to the propagator(s), thus giving
finite functions $f_i^g(x,y,\theta_1,\theta_2)$ which can safely be
implemented in a computer program.  The modified phase space
$\overline{d\Gamma}_3$
is expanded using generalized plus distributions allowing extraction
of soft and/or collinear poles in $n-4$ which are
cancelled as usual.

In this particular case, as the photon is virtual, and the heavy quarks
are massive, the only collinear singularity comes from the collinear
emission of a gluon from the incoming gluon as shown in fig.\ 3(d).
The squared matrix element $M_i^g(3)$ has a \mbox{$1 / (1+y)$} singularity
when \mbox{$y \rightarrow -1$}, therefore it suffices to multiply by
\mbox{$t^{\prime} \propto (1+y)(1-x)$} to get a result that is finite as
\mbox{$y \rightarrow -1$}.  Similarly, the emission of a
soft gluon corresponds to a $1/(1-x)^2$ singularity in $M_i^g(3)$ as
\mbox{$x \rightarrow 1$} so we choose to multiply by an additional factor of
\mbox{$( u^{\prime}-q^2 s_5/s) \propto (1-x)$} to get a result that
is finite as \mbox{$x \rightarrow 1$}.

We now proceed to show the cancellation of the singularities and
derive the renormalized partonic cross sections.
Substituting the expressions for $t^{\prime}$ and $u^{\prime}$ found
in appendix A and using eq.\ (\ref{ps3}) we find
\begin{equation}
\overline{d\Gamma_3} = \frac{2}{\pi} H N d\Gamma_2^{(5)}
	(s^{\prime})^{-1+\epsilon/2} \left(
	\frac{s^{\prime}}{s} \right)^{-1+\epsilon/2} (1-x)^{-1+\epsilon}
	(1-y^2)^{-1+\epsilon/2} dy \sin^{\epsilon} \theta_2 d \theta_2 \, ,
\end{equation}
where all quantities are defined in appendix A.  We now begin to
extract the poles,
by expanding the $(1-x)^{-1+\epsilon}$ piece in terms of the generalized
plus distributions defined in appendix A. Using eq.\ (\ref{xplus}) one finds
\begin{eqnarray}
d \sigma_{i,g}^{(3)} & = & \frac{2}{\pi}C_{i,g}HNd\Gamma_2^{(5)}
(s^{\prime})^{-1+\epsilon/2} \left( \frac{s^{\prime}}{s}
\right)^{-1+\epsilon/2} \delta (1-x) \left[ \frac{1}{\epsilon}
+ 2 \ln \tilde{\beta}
+ 2 \epsilon \ln^2 \tilde{\beta} \right] \nonumber \\
& \times & (1-y^2)^{-1+\epsilon/2}
dy \sin^{\epsilon} \theta_2 d\theta_2 f_i^g(x,y,\theta_1,\theta_2)
\nonumber \\
&+& \frac{2}{\pi}C_{i,g}HNd\Gamma_2^{(5)}
(s^{\prime})^{-1+\epsilon/2} \left(
\frac{s^{\prime}}{s} \right)^{-1+\epsilon/2}
\left[ \left( \frac{1}{1-x} \right)_{\tilde{\rho}} + \epsilon
\left( \frac{\ln(1-x)}{1-x} \right)_{\tilde{\rho}} \right] \nonumber \\
& \times &
(1-y^2)^{-1+\epsilon/2}
dy \sin^{\epsilon} \theta_2 d\theta_2 f_i^g(x,y,\theta_1,\theta_2).
\end{eqnarray}
The $\delta(1-x)/ \epsilon$ piece in the first term is a soft
singularity now regulated dimensionally.  In the first piece we use
$d \Gamma_2^{(5)} \delta(1-x) = \delta(1-x) d \Gamma_2 dx$ while
in the second piece we expand
$(1-y^2)^{-1+\epsilon/2}$ using eq.\ (\ref{yplus2}) to find,
\begin{equation}
d\sigma_{i,g}^{(3)}=d\sigma_{i,g}^{(s)}+d\sigma_{i,g}^{(c-)}
+d\sigma_{i,g}^{(f)},
\end{equation}
where
\begin{equation}
\label{gs}
d\sigma_{i,g}^{(s)} = \frac{2}{\pi}C_{i,g}HNd\Gamma_2
	(s^{\prime})^{-1+\epsilon/2} \left(
	\frac{s^{\prime}}{s} \right)^{-1+\epsilon/2}
\left[ \frac{1}{\epsilon}+2\ln\tilde{\beta}+2\epsilon\ln^2\tilde{\beta} \right]
f^g_i(\theta_1),
\end{equation}
with
\begin{equation}
f^g_i(\theta_1) = \int \delta(1-x)dx(1-y^2)^{-1+\epsilon/2}dy \sin^{\epsilon}
\theta_2 d\theta_2 f^g_i(x,y,\theta_1,\theta_2),
\end{equation}
and
\begin{eqnarray}
\label{gc}
d\sigma_{i,g}^{(c-)} & = & \frac{2}{\pi}C_{i,g}HNd\Gamma_2^{(5)}
(s^{\prime})^{-1+\epsilon/2} \left(
\frac{s^{\prime}}{s} \right)^{-1+\epsilon/2}
\left[ \left( \frac{1}{1-x} \right)_{\tilde{\rho}} + \epsilon
\left( \frac{\ln(1-x)}{1-x} \right)_{\tilde{\rho}} \right] \nonumber \\
& \times &
\delta(1+y) \left(\frac{1}{\epsilon} + \frac{1}{2} \ln 2 \omega \right)
dy \sin^{\epsilon} \theta_2 d\theta_2 f_i^g(x,y,\theta_1,\theta_2),
\end{eqnarray}
and
\begin{eqnarray}
\label{gf}
d\sigma_{i,g}^{(f)} & = & \frac{1}{\pi}C_{i,g}HNd\Gamma_2^{(5)}
(s^{\prime})^{-1+\epsilon/2} \left(
\frac{s^{\prime}}{s} \right)^{-1+\epsilon/2}
\left[ \left( \frac{1}{1-x} \right)_{\tilde{\rho}} + \epsilon
\left( \frac{\ln(1-x)}{1-x} \right)_{\tilde{\rho}} \right] \nonumber \\
& \times &
\left[ \left( \frac{1}{1+y} \right)_{\omega}
+ \left( \frac{1}{1-y} \right)_{\omega} \right]
dy \sin^{\epsilon} \theta_2 d\theta_2 f_i^g(x,y,\theta_1,\theta_2).
\end{eqnarray}

According to the factorization theorem \cite{CSS} the cross section
for scattering of a virtual-photon off a hadron $H$ of momentum $p$ is
\begin{equation}
\label{Hfact}
d \sigma_H(p) = \sum_i \int_{0}^{1} d \hat{\sigma}_i (\xi p)
\phi_{i/H}(\xi,\mu^2_f) d \xi,
\end{equation}
where the sum runs over all partons in $H$ and
we have suppressed the polarization index temporarily.
Here $\phi_{i/H}(\xi,\mu^2_f)$ is the probability density for finding
parton $i$ in hadron $H$ with momentum fraction between $\xi$ and
$\xi + d \xi$ at scale $\mu_f$.
Noting that the
infrared safe $d \hat{\sigma}_i$ is independent of the external hadron,
we are free to set $H=j$, where $j$ represents a parton.  Then the
partonic cross sections satisfy
\begin{equation}
\label{pfact}
d \sigma_j (k_1) = \sum_i \int_{0}^{1} d \hat{\sigma}_{i} (xk_1)
\phi_{i/j}(x,\mu^2_f) dx.
\end{equation}
Up to first order in $\alpha_s$
\begin{equation}
\label{split}
\phi_{i/j}(x,\mu^2_f) = \delta_{ij} \delta (1-x)
+ \frac{\alpha_s}{2 \pi} \left[
P_{ij}(x) \frac{2}{\epsilon} + f_{ij}(x,\mu^2_f) \right],
\end{equation}
where $P_{ij}(x)$ denote the Altarelli-Parisi splitting functions \cite{AP}
and the functions $f_{ij}(x,\mu_f^2)$ depend on the mass factorization
scheme chosen.  In the $\overline{{\rm MS}}$ scheme
\begin{equation}
\label{f}
f_{ij}(x,\mu^2_f) = P_{ij}(x) \left( \gamma_{E} - \ln 4 \pi + \ln \mu_f^2/
\mu^2 \right),
\end{equation}
where $\gamma_E$ is the Euler-Mascheroni constant.
In this paper we choose the mass factorization scale $\mu_f$
to be equal to the renormalization scale $\mu$ so that the last term in
eq.\ (\ref{f}) is zero.  Expanding $d \sigma_j$ and $d \hat{\sigma}_i$
in powers of $\alpha_s$ gives
\begin{equation}
\label{series1}
d \sigma_j (k_1) = d \sigma_j^{(0)} (k_1) + d \sigma_j^{(1)}  (k_1) +
	\cdots
\end{equation}
\begin{equation}
\label{series2}
d \hat{\sigma}_i (xk_1) = d \hat{\sigma}_i^{(0)} (xk_1) +
	d \hat{\sigma}_i^{(1)} (xk_1) + \cdots .
\end{equation}
Substituting eqs.\ (\ref{split}), (\ref{f}), (\ref{series1}),
and (\ref{series2}) into eq.\ (\ref{pfact}) and equating powers of
$\alpha_s$ we obtain
\begin{eqnarray}
\label{sig0}
d \hat{\sigma}_j^{(0)} (k_1) & = & d \sigma_j^{(0)} (k_1), \\
\label{sig1}
d \hat{\sigma}_j^{(1)} (k_1) & = &  d \sigma_j^{(1)}  (k_1) -
	\frac{\alpha_s}{2 \pi} \sum_i \int_0^1 dx \, d \sigma_i^{(0)}
	(xk_1) P_{ij}(x) \frac{2}{\bar{\epsilon}},
\end{eqnarray}
where $2/ \overline{\epsilon} = 2/ \epsilon + \gamma_E - \ln 4 \pi$.

In the collinear limit $y \rightarrow -1$ we show in appendix B that
\begin{equation}
\label{gsep}
f^g_i(x,-1,\theta_1,\theta_2) = f^g_i(x,\theta_1)
+ \tilde{f}^g_i(x,\theta_1,\theta_2),
\end{equation}
where
\begin{equation}
\label{gint}
\int_0^{\pi} \tilde{f}^g_i(x,\theta_1,\theta_2) \sin^{\epsilon} \theta_2
d \theta_2 = 0,
\end{equation}
and
\begin{eqnarray}
\label{gfxtheta1}
f^g_i(x,\theta_1) &=& 512 \pi^2 \mu^{-2 \epsilon} \alpha_s^2 e^2 e^2_H N
C_A C_F \frac{(s^{\prime})^2}{s} \nonumber \\ &\times&
\frac{1-x}{x} \left[
x(1-x) + \frac{x}{1-x} + \frac{1-x}{x} \right] B_{i,QED}(xk_1) .
\end{eqnarray}
Performing the $y$ and $\theta_2$ integrations in eq.\ (\ref{gc}) using
eqs.\ (\ref{gsep}), (\ref{gint}), and (\ref{gfxtheta1}) we find
\begin{eqnarray}
\label{sigc1}
d \sigma_{i,g}^{(c-)} &=& 32 C_{i,g} \mu^{- \epsilon}
\alpha_s^2 e^2 e^2_H N C_A C_F
d \Gamma^{(5)}_2 \nonumber \\ & \times &
\left\{ \frac{2}{\bar{\epsilon}} \left( \frac{1}{1-x} \right)_{\tilde{\rho}}
+ 2 \left( \frac{\ln(1-x)}{1-x} \right)_{\tilde{\rho}}
+ \left( \frac{1}{1-x} \right)_{\tilde{\rho}} \left[
\ln \frac{s^{\prime}}{\mu^2} + \ln \frac{s^{\prime}}{s}
+  \ln \frac{\omega}{2} \right]
\right\} \nonumber \\ & \times &
\frac{1-x}{x}
\left[ x(1-x) + \frac{x}{1-x} + \frac{1-x}{x} \right] B_{i,QED}(xk_1) .
\end{eqnarray}
Appealing to eq.\ (\ref{sig1}) and reinstating the polarization
index $i$  we find
\begin{eqnarray}
\label{factsum}
d \hat{\sigma}_{i,g}^{(v)} + d \hat{\sigma}_{i,g}^{(s)}
+  d \hat{\sigma}_{i,g}^{(c-)} &=& d \sigma_{i,g}^{(v)} +
d \sigma_{i,g}^{(s)} + d \sigma_{i,g}^{(c-)}
\nonumber \\
&-& \frac{\alpha_s}{2 \pi} \int dx d \sigma_{i,g}^{(0)} (xk_1) P_{gg}(x)
\frac{2}{\bar{\epsilon}},
\end{eqnarray}
where we have used $d \sigma_{i,q}^{(0)}=d \sigma_{i,\bar{q}}^{(0)}=0$.
The Born level cross sections $d \sigma_{i,g}^{(0)}$ are given
by eqs.\ (\ref{sigj}) and (\ref{Mborn}) and the virtual corrections denoted
by $d \sigma_{i,g}^{(v)}$ are discussed below.  It is well known that \cite{AP}
\begin{eqnarray}
\label{Pgg}
P_{gg}(x) &=& 2C_A \left[ \frac{x}{(1-x)_{+}} + \frac{1-x}{x}+x(1-x) \right]
+ \left( \frac{11}{6} C_A - \frac{2}{3} T_F n_{lf} \right) \delta(1-x) ,
\nonumber \\ &=&
2C_A \left[ \frac{x}{(1-x)_{\tilde{\rho}}} + \frac{1-x}{x}+x(1-x) \right]
\nonumber \\
&+& \left( \frac{11}{6} C_A - \frac{2}{3} T_F n_{lf} +4C_A \ln \tilde{\beta}
\right) \delta(1-x), \nonumber \\ &&
\end{eqnarray}
where $T_F=1/2$, and $n_{lf}$ is the number of light quarks.  In the second
line we have exchanged the usual plus distribution for
the more general $\tilde{\rho}$ distribution and defined
$\tilde{\beta} = \sqrt{1-\tilde{\rho}}$.
Cancelling the pole in eq.\ (\ref{sigc1}) by using eq.\ (\ref{factsum})
gives the ``factorized collinear'' term
\begin{eqnarray}
d \hat{\sigma}_{i,g}^{(c-)} & =& 32 C_{i,g} \alpha_s^2 e^2 e_H^2 N C_A C_F
	B_{i,QED}(xk_1) \, d \Gamma_2^{(5)} \! \left[ (1-x)^2 + 1
	+ \frac{(1-x)^2}{x^2} \right] \nonumber \\
& &\times \left[ \left( \frac{1}{1-x} \right)_{\tilde{\rho}}
	\left( \ln \frac{s^{\prime}}{\mu^2}
	+ \ln \frac{s^{\prime}}{s} + \ln \frac{\omega}{2} \right)
	+ 2 \left( \frac{\ln(1-x)}{1-x} \right)_{\tilde{\rho}} \right] \,.
\end{eqnarray}
In the limit $Q^2=0$ this result reduces to formula (2.45)
of \cite{frix} when one identifies
$M^{(b)}_{\gamma g} = 32 C_{i,g} \pi \alpha_s e^2 e^2_H N C_F B_{G,QED}$.
The remaining poles in eq.\ (\ref{factsum}) cancel when
the soft and virtual cross sections are added according to
\begin{eqnarray}
\label{factsumf}
d \hat{\sigma}_{i,g}^{(s+v)} &=& d \sigma_{i,g}^{(v)} + d \sigma_{i,g}^{(s)}
\nonumber \\
&-& 16 C_{i,g} \alpha_s^2 \mu^{- \epsilon} e^2 e^2_H N C_F d \Gamma_2
\frac{2}{\bar{\epsilon}}
\left( \frac{11}{6} C_A - \frac{2}{3} T_F n_{lf} + 4C_A \ln \tilde{\beta}
\right) B_{i,QED} (k_1). \nonumber \\ &&
\end{eqnarray}
The soft cross sections $d \sigma_{i,g}^{(s)}$, which follow from
eq.\ (\ref{gs}), are given in appendix B\@.
The one loop order $eg^3$ diagrams for the virtual corrections
to the reaction (\ref{born}) are shown in fig.\ 4.
The interference of these diagrams with the lowest order diagrams
of fig.\ 2 were calculated in \cite{LRSvN1} using the renormalization
scheme of \cite{JC} and are available as FORM code.
Using (\ref{factsumf}) we analytically checked the cancellation of
the singularities.  Also available from \cite{LRSvN1}
as FORTRAN code is the analog of eq.\ (\ref{factsumf}) using the
$\Delta$ prescription \cite{delta} for dividing the phase space into soft
and collinear regions.  The components of (\ref{factsumf}) in the
$\Delta$ prescription and the subtraction method differ by finite terms.
Therefore, we modified the original FORTRAN code by adding and subtracting
appropriate finite pieces.  In fact, the soft finite pieces to subtract
off are given in eqs.\ (3.24) and (3.25)
of \cite{LRSvN1}, and the pieces to add on are
the finite pieces of eqs.\ (\ref{SOK}) and (\ref{SQED}) of appendix B.
Note that there is a typographical error in eq.\ (3.25) of \cite{LRSvN1};
the sign of the $\ln^2 r_s$ term should be negative not positive.

Finally the three body contributions $d \sigma_{i,g}^{(f)}$ are
finite in $n=4$ space-time dimensions and a short calculation from (\ref{gf})
shows they may be written as
\begin{eqnarray}
d \sigma_{i,g}^f & = & 2\left( \frac{1}{16 \pi^2} \right)^2 C_{i,g} \beta_5
	\frac{s}{(s^{\prime})^2} \left( \frac{1}{1-x} \right)_{\tilde{\rho}}
	\left( \frac{1}{1+y} \right)_{\omega} \frac{1}{1-y} \nonumber \\
& &\times f^g_i (x,y,\theta_1,\theta_2) dx dy \sin \theta_1 d
	\theta_1 d \theta_2 \, ,
\end{eqnarray}
where $\beta_5 = \sqrt{1-4m^2/s_5}$.  The finite functions
$f^g_i (x,y,\theta_1,\theta_2)$ follow from
the definition (\ref{fgi}), eq.\ (\ref{M}), and
the diagrams in fig.\ 3.  Summarizing, we have the final result
\begin{equation}
\label{siggf}
d \hat{\sigma}_{i,g} = d \sigma_{i,g}^{(0)} + d \hat{\sigma}_{i,g}^{(s+v)}
	+ d \hat{\sigma}_{i,g}^{(c-)} + d \sigma_{i,g}^{(f)} \, ,
\end{equation}
that is finite in $n=4$ space-time dimensions and is to be used in
eq.\ (\ref{Hfact}) to make predictions.  Each of the last three terms in
(\ref{siggf}) individually depends on $\tilde{\rho}$ and $\omega$.  However,
the sum is independent of them as it must be because the decomposition
(\ref{siggf}) follows directly from eqs.\ (\ref{yplus2}) and (\ref{xplus}).
This provides a strong check on our computer code.

\subsection{$\gamma^{\ast}q$ and $\gamma^{\ast}\bar q$ Channels}

Analysis of the partonic reaction(s)
\begin{equation}
\label{quark}
\gamma^{\ast}(q) + q(k_1)(\bar{q}(k_1)) \rightarrow
q(k_2)(\bar{q}(k_2))+Q(p_1)+\bar{Q}(p_2) \, ,
\end{equation}
proceeds as above only this time there are no soft or virtual
contributions making the analysis much simpler.
This time we write the partonic cross sections as
\begin{eqnarray}
d \sigma_{i,q}^{(3)} &=& C_{i,q} M^q_i (3) d \Gamma_3 \nonumber \\
&=& C_{i,q} f^q_i(x,y,\theta_1,\theta_2) \overline{d\Gamma_3} \, ,
\end{eqnarray}
where
\begin{equation}
\label{fqi}
f^q_i(x,y,\theta_1,\theta_2) \equiv t^{\prime} \, M^q_i(3) \, ,
\end{equation}
and
\begin{equation}
\overline{d\Gamma_3} \equiv d\Gamma_3 / t^{\prime}.
\end{equation}
Replacing the divergent factor $(1+y)^{-1+\epsilon/2}$ in
$\overline{d\Gamma_3}$ by eq.\ (\ref{yplus1}) one obtains the
following decomposition:
\begin{equation}
d \sigma_{i,q}^{(3)}=d \sigma^{(c-)}_{i,q}+d \sigma^{(f)}_{i,q},
\end{equation}
with
\begin{eqnarray}
d \sigma^{(c-)}_{i,q} &=& - \frac{1}{\pi} C_{i,q} H N d\Gamma_2^{(5)}
(s^{\prime})^{\epsilon/2} \left( \frac{s^{\prime}}{s} \right)^{\epsilon/2}
(1-x)^{\epsilon} (1-y)^{\epsilon/2}
\nonumber \\ & \times &
\delta (1+y) \left( \frac{2}{\epsilon} + \ln \omega \right)
dy \sin^{\epsilon} \theta_2 d\theta_2 f_i^q(x,y,\theta_1,\theta_2),
\end{eqnarray}
and
\begin{eqnarray}
d \sigma^{(f)}_{i,q} &=& - \frac{1}{\pi} C_{i,q} H N d\Gamma_2^{(5)}
(s^{\prime})^{\epsilon/2} \left( \frac{s^{\prime}}{s} \right)^{\epsilon/2}
(1-x)^{\epsilon} (1-y)^{\epsilon/2}
\nonumber \\ & \times &
\left( \frac{1}{1+y} \right)_{\omega}
dy \sin^{\epsilon} \theta_2 d\theta_2 f_i^q(x,y,\theta_1,\theta_2).
\end{eqnarray}
As in the gluon-bremsstrahlung reaction (see appendix C)
\begin{equation}
f^q_i(x,-1,\theta_1,\theta_2)=f^q_i(x,\theta_1)
+\tilde{f}^q_i(x,\theta_1,\theta_2) \, ,
\end{equation}
with
\begin{equation}
\int_0^{\pi} \tilde{f}^q_i(x,\theta_1,\theta_2) \sin^{\epsilon}
\theta_2 d \theta_2 = 0.
\end{equation}
In this case we find that
\begin{eqnarray}
\label{fqint}
f^q_i(x,\theta_1) &=& -128 \pi^2 \mu^{-2 \epsilon}
\alpha_s^2 e^2 e^2_H N C_F (1+\epsilon/2)^{-1} \nonumber \\ & \times &
\left[ \frac{1+(1-x)^2+ \epsilon x^2/2}{x^2} \right] B_{i,QED} (xk_1).
\end{eqnarray}
Appealing to eq.\ (\ref{sig1}) with \cite{AP}
\begin{equation}
P_{gq}(x) = C_F \left[ \frac{1+(1-x)^2}{x} \right],
\end{equation}
one finds
\begin{equation}
\label{quarkfinal}
d \hat{\sigma}_{i,q}^{(3)}=d\hat{\sigma}^{(c-)}_{i,q}+d\sigma^{(f)}_{i,q},
\end{equation}
with
\begin{eqnarray}
d \hat{\sigma}^{(c-)}_{i,q} &=& 8 C_{i,q} \alpha^2_s e^2 e_H^2 N C_F
B_{i,QED}(xk_1) d \Gamma_{2}^{(5)} \nonumber \\ & \times &
\left\{ 1+ \frac{1+(1-x)^2}{x^2}
\left[ \ln\frac{s^{\prime}}{\mu^2}
+ \ln\frac{s^{\prime}}{s} + \ln \frac{\omega}{2} + 2\ln(1-x)
\right] \right\} \, , \\
d\sigma^{(f)}_{i,q} & = & - \left( \frac{1}{16 \pi^2} \right)^2 C_{i,q}
\beta_5 f^q_i(x,y,\theta_1,\theta_2) \left( \frac{1}{1+y} \right)_{\omega}
dx dy \sin \theta_1 d\theta_1d\theta_2. \nonumber \\
&&
\end{eqnarray}
The finite functions $f^q_i (x,y,\theta_1,\theta_2)$ follow from
the definition (\ref{fqi}), eq.\ (\ref{M}), and
the diagrams in fig.\ 5.
In the limit $Q^2=0$, the ``factorized collinear'' cross sections for the
quark channel $d \hat{\sigma}^{(c-)}_{i,q}$ reduces to formula (2.62)
of \cite{frix} when one identifies
$M^{(b)}_{\gamma g} = 32 C_{i,g} \pi \alpha_s e^2 e^2_H N C_F B_{G,QED}$ and
uses eq.\ (\ref{Crelation}).
This completes the analysis of the quark induced reaction.  The
analysis of the antiquark induced reaction is completely analogous.

As the quark channel only contains collinear singularities,
we will use it to illustrate how the generalized plus
distributions are implemented numerically
and how the $\omega$ dependence disappears in the sum (\ref{quarkfinal}).
To this end consider
\begin{eqnarray}
\label{distex}
d\sigma^{(f)}_{i,q} & \sim & \int_{-1}^1 dy f(y) \left( \frac{1}{1+y}
\right)_{\omega} \nonumber \\\
& = & \int_{-1}^{-1+\omega} dy f(y) \left( \frac{1}{1+y} \right)_{\omega}
  +   \int_{-1+\omega}^1 dy f(y) \left( \frac{1}{1+y} \right)_{\omega}
\nonumber \\
& = & \int_{-1}^{-1+\omega} dy \frac{f(y)-f(-1)}{1+y}
  +   \int_{-1+\omega}^1 dy \frac{f(y)}{1+y}
\nonumber \\
& = & \int_{-1}^1 dy \frac{f(y)}{1+y}
  -   \int_{-1}^{-1+\omega} dy \frac{f(-1)}{1+y}
\end{eqnarray}
where we have suppressed all indices and arguments of
$f^q_i(x,y,\theta_1,\theta_2)$ other than $y$.
In the bottom line we see that the infinity encountered at the lower
integration limit $y=-1$ is cancelled in the sum of the two integrals.
In practice one can only reasonably take the lower limit to be
$-1+\delta$ where $\delta \sim 10^{-7}$ in double
precision FORTRAN before round off errors enter.  None the less, the final
results are stable with respect to the variation of $\delta$ in the
range $10^{-5}$ to $10^{-7}$.  The upper
limit of the second integral gives a contribution $f(-1) \ln \omega$ which
cancels against the $\ln \omega$ term in $d \hat{\sigma}^{(c-)}_{i,q}$.
The first integral in the bottom line is commonly called the ``event'' and
has a positive definite weight.  The second integral plus the factorized
collinear contribution is commonly called the ``counter-event'' and may have
either positive or negative weight.  The implementation for the gluon channel
is similar but more complicated due to the presence of the soft-virtual terms.

\mysection{Results}

\subsection{Virtual-photon-parton cross sections}

In this section we discuss the virtual-photon-parton cross sections and
compare them with the results of \cite{LRSvN1}.  The total cross sections
are obtained by integration so that
\begin{equation}
\hat{\sigma}_{k,i} = \int d \hat{\sigma}_{k,i} \, ,
\end{equation}
where $d \hat{\sigma}_{k,i}$ are given as eqs.\ (\ref{siggf}), and
(\ref{quarkfinal}).
Choosing a renormalization scheme where the heavy quarks in the gluon
self-energy loops decouple in the limit of small momenta
flowing into the loop \cite{JC}, we express the perturbative expansion of the
virtual photon-parton cross section in terms of scaling functions as follows:
\begin{equation}
\label{scalef}
\hat{\sigma}_{k,i}(s,q^2,m^2)
= \frac{\alpha \alpha_s}{m^2} \left[ f^{(0)}_{k,i}
( \eta, \xi ) + 4 \pi \alpha_s \left\{ f^{(1)}_{k,i} ( \eta, \xi ) +
\bar{f}^{(1)}_{k,i} ( \eta, \xi ) \ln \frac{\mu^2_f}{m^2} \right\} \right],
\end{equation}
where
\begin{equation}
\eta = \frac{s}{4m^2}-1, \quad \xi = \frac{Q^2}{m^2},
\end{equation}
with $s$ the square of the c.\ m.\
energy in the virtual-photon-parton system and $\mu_f^2$ the mass
factorization scale.
Since the $f^{(l)}_{k,i}( \eta, \xi )$ depend on the electric charges of the
heavy and/or light quarks we extract these charges and define
new functions via
\begin{eqnarray}
\label{fcharge}
f^{(l)}_{k,g}( \eta, \xi ) & = & e_H^2 c^{(l)}_{k,g}( \eta, \xi ), \nonumber \\
f^{(1)}_{k,q}( \eta, \xi ) & = & e_H^2 c^{(1)}_{k,q}( \eta, \xi ) +
e_H e_L o^{(1)}_{k,q}( \eta, \xi ) + e_L^2 d^{(1)}_{k,q}( \eta, \xi ) \, ,
\end{eqnarray}
together with the corresponding formulae for the coefficient of the mass
factorization term.  Note that the $o^{(1)}_{k,q}( \eta, \xi )$ vanish
after integration to form the total photon-parton cross sections
so they were not needed in \cite{LRSvN1}.  However, now they must be retained
when plotting distributions.  Further, $d^{(1)}_{k,q}( \eta, \xi )$
can only be
evaluated at $\xi=0$ provided the additional collinear divergence which
arises when the photon goes on-mass-shell is mass factorized.  As we only
consider $Q^2 \geq 8.5 ( {\rm GeV/c} )^2$ we do not perform this
factorization although it was performed in \cite{LRSvN1} and checked
against the photoproduction limit in \cite{SvN}.
$m=4.75$ (GeV/$c^2$) {\em i.e.} the corrected version of fig.\ 11(b) of

Upon integration of the eqs.\ (\ref{siggf}), and (\ref{quarkfinal}) we
find agreement with all the plots shown in \cite{LRSvN1} except for
discrepancies with three figures, namely, fig.\ 9(b), fig.\ 11(a),
and fig.\ 11(b).  The correct plots are reproduced here as
fig.\ 6, fig.\ 7(a), and fig.\ 7(b), respectively.
This checking also uncovered minor errors in the previous FORTRAN
programs for the inclusive calculations which have now been corrected.
Fortunately, they were all in the virtual-photon-quark channels so were
not significant numerically and do not alter any of the other plots or
results in references \cite{LRSvN1}, \cite{LRSvN2}, and \cite{RSvN}.

\subsection{Hadronic structure functions and correlations}

Recalling that the probability density is related to the momentum
density via $f_{i/H} (\xi,\mu^2_f) = \xi \phi_{i/H}(\xi,\mu^2_f)$ we
write (\ref{Hfact}) as
\begin{equation}
d \sigma_{\gamma^{\ast}H}(p) = \sum_i \int_{0}^{1} \frac{d \xi}{\xi}
d \hat{\sigma}_i (\xi p) f_{i/H}(\xi,\mu^2_f).
\end{equation}
We now specialize to the case where $H$ is a proton, as in the case of
HERA.  Using the relations
\begin{equation}
F_k = \frac{Q^2}{4 \pi^2 \alpha} \sigma_k,
\end{equation}
where $k=2,L$ with $\sigma_2 = \sigma_G + 3 \sigma_L / 2$, and the relations
for the scaling functions (\ref{scalef}), and (\ref{fcharge}), we find
\begin{eqnarray}
F_{k}(x,Q^2,m^2) &=&
\frac{Q^2 \alpha_s(\mu^2)}{4\pi^2 m^2}
\int_{\xi_{\rm min}}^1 \frac{d\xi}{\xi}  \Big[ \,e_H^2 f_{g/P}(\xi,\mu^2)
 c^{(0)}_{k,g} \,\Big] \nonumber \\&&
+\frac{Q^2 \alpha_s^2(\mu^2)}{\pi m^2}
\int_{\xi_{\rm min}}^1 \frac{d\xi}{\xi} \left\{ \,e_H^2 f_{g/P}(\xi,\mu^2)
\left( c^{(1)}_{k,g} + \bar c^{(1)}_{k,g} \ln \frac{\mu^2}{m^2} \right) \right.
\nonumber \\ &&
+\sum_{i=q,\bar q} f_{i/P}(\xi,\mu^2) \left. \left[ e_H^2
\left( c^{(1)}_{k,i} + \bar c^{(1)}_{k,i} \ln \frac{\mu^2}{m^2} \right)
+ e^2_i \, d^{(1)}_{k,i} + e_i\, e_H \, o^{(1)}_{k,i} \, \right]
\right\} \, , \nonumber \\ &&
\end{eqnarray}
where $k = 2,L$.  We have set $\mu_f=\mu$ and shown the $\mu^2$
dependence of the running
coupling $\alpha_s$ explicitly.  The lower boundary on the integration
is given by $\xi_{\rm min} = x(4m^2+Q^2)/Q^2$.
This formula yields the standard heavy quark
structure functions $F_2(x,Q^2,m^2)$ and $F_L(x,Q^2,m^2)$ for
electron proton scattering, and we will present results as differentials
of these functions.

{}From the formalism described in the previous section we are left with
events of positive weight and counter-events of either positive or
negative weight.  Our program
outputs the final state four vectors of the event and counter-event
together with the corresponding weight.
We histogram these into bins to produce differential distributions.

We note that the inclusive distributions in the
heavy quark transverse momentum,
( i.e., $dF_2(x,Q^2,m^2,p_t)/dp_t$ and $dF_L(x,Q^2,m^2,p_t)/dp_t$)
and in the heavy quark rapidity
( i.e., $dF_2(x,Q^2,m^2,y)/dy$ and $dF_L(x,Q^2,m^2,y)/dy$)
were already published in \cite{LRSvN2}
for a range of $x$ values at fixed $Q^2$ and for a range of $Q^2$
values at fixed $x$.  These ranges are covered by the H1 and ZEUS
detectors in HERA.  We have reproduced these plots with our new programs.
Also, we have previously published distributions in the
invariant mass of the $Q \bar{Q}$ pair,
( i.e., $dF_2(x,Q^2,m^2,M)/dM$ and
$dF_L(x,Q^2,m^2,M)/dM$) for charm production in \cite{HS}.
Therefore we concentrate on new results involving
distributions which are sensitive to the four momenta
of all the final particles.

We start by considering the production of charm quarks.  We use
$m=m_c=1.5 \, {\rm GeV/c^2}$ and simply choose the factorization
(renormalization) scale as $\mu^2=Q^2+4(m_c^2+( P_t^c + P_t^{\bar{c}} )^2/4)$.
Note that there are many possible choices of scale as we have all
components of the final four vectors.  Aside from the $P_t$ dependence,
this choice reduces to the
usual choice of $\mu^2=Q^2$ for electroproduction of massless quarks and
$\mu^2=4m_c^2$ for the photoproduction of charm quarks.  We introduce a $P_t$
dependence by adding in the average of the magnitude of the
transverse momenta of the heavy quark and heavy antiquark.
As mentioned earlier we use the CTEQ3M parton densities \cite{CTEQ} in the
$\overline{\rm MS}$ scheme and the two loop $\alpha_s$ with
$\Lambda_{4} = 0.239 \, {\rm GeV}$.

The first distribution we present basically measures the transverse momentum
of the additional jet which recoils against the heavy quark pair.
The $P_t$ distribution of the charm-anticharm pair is shown
in fig.\ 8(a)
where we plot $dF_2(x,Q^2,m_c^2,P_t) / dP_t$
as a function of $P_t$.
The histograms are presented at fixed $Q^2$= 12 $({\rm  GeV}/c)^2$
for $x$ values of $4.2 \times  10^{-4}$, $8.5 \times  10^{-4}$,
$1.6 \times  10^{-3}$ and $2.7 \times  10^{-3}$ respectively.
One sees that the $P_t$ distribution peaks at small
$P_t$ and
has a small negative contribution in the lowest bin.
This is a region where the dominant contribution is from counter-events
so the weights can be negative.
The results of this calculation require missing contributions
from even higher order perturbation theory before this bin will have a
positive weight.
In general there will be corresponding bins in all the exclusive plots
where the weights can be negative.  The depth of the negative bins
is a function of $x$, $Q^2$, and the choice of scale.
Note that at larger $P_t$ the structure
function is dominated by the contribution from the square of the
bremsstrahlung graphs so the weights are positive.
Figure 8(b) shows the corresponding results for fixed $x= 8.5 \times 10^{-4}$
plotted for the $Q^2$ values of
$8.5$ $({\rm GeV}/c)^2$, $12$ $({\rm GeV}/c)^2$,
$25$ $({\rm GeV}/c)^2$ and $50$ $({\rm GeV}/c)^2$ respectively.
The distributions peak near small $P_t$ and are either small
or negative in the first bin.  The histograms with
$Q^2 = 12 ( {\rm GeV} / c )^2 $ and $x=8.5 \times 10^{-4}$
(the dotted line) are the same in figs.\ 8(a) and 8(b).  We have also
used the same scales on the axes so one can easily see that there is a greater
change if we fix $x$ and vary $Q^2$ than if we fix $Q^2$ and vary $x$.  We
will continue to use the same scale for all the pairs of later plots to
simplify the comparison between them.

We now turn to the distributions in the azimuthal angle between the
outgoing charm quark and charm antiquark which we denote as $\Delta \phi$.
This is the angle between the ${\bf P}_t$ vectors of the heavy quark-antiquark
in the c.\ m.\ frame of the virtual-photon-hadron system.
Since we integrate over the azimuthal
angle between the plane containing the incoming and outgoing leptons and the
plane defined by the incoming parton and outgoing heavy quark
(to limit our discussion to $F_2$ and $F_L$) we can only
plot relative azimuthal correlations.
In the Born approximation this distribution is a delta function at $\pi$,
as their four momenta must balance.  Due to the radiation of the additional
light mass parton, the distribution has a tail extending below $\pi$
and has a valley at $\pi$.
The distributions become negative in the
highest bins. This negative region is a general feature of all
exclusive calculations.
Figure 9(a) shows results for $dF_2(x,Q^2,m_c^2,\Delta \phi) / d(\Delta \phi)$
at the same values of fixed $Q^2$ and variable $x$ as chosen
previously in fig.\ 8(a), while fig.\ 9(b) shows the results for fixed $x$
and variable $Q^2$ as chosen previously in fig.\ 8(b).  Note again
that the dotted histograms are the same in fig.\ 9(a) and 9(b), and
there is more variation for fixed $x$ and changing $Q^2$ than for
fixed $Q^2$ and changing $x$.

Finally we show distributions in the so-called heavy quark cone size variable
$R = \sqrt{(\Delta \phi)^2 + (\Delta \eta)^2}$ where
$\Delta \phi$ is the azimuthal angle between the ${\bf P}_t$ vectors of
charm-anticharm quarks in the c.\ m.\ frame of the virtual-photon-hadron
system and $\Delta \eta$ is the difference in
pseudo-rapidities of the charm quark-antiquark pair.  We define
pseudo-rapidity to be
\mbox{$\eta = 1 / 2 \ln \left[ (1+\cos \theta) / (1-\cos \theta ) \right]$}
where $\theta$ is the angle the quark (antiquark) makes with the
axis defined in the back-to-back photon-hadron system.
In fig.\ 10(a) we choose fixed $Q^2$ and variable $x$
as in fig.\ 8(a) and plot $dF_2(x,Q^2,m_c^2,R) / dR$.
In fig.\ 10(b) we choose fixed $x$ and variable $Q^2$ as in fig.\ 8(b).
The influence of the additional radiation causes these
distributions to peak below $R = \pi$ and develop a dip
at values above $R=\pi$. The negative bins are unphysical
and should be partially filled in by higher order corrections.  The
depth of the negative bins is a function of the bin width.  If we use a
wider bin, the dips will be less pronounced.  Again, as in the previous
cases, we note there is a greater change in the histograms when we fix $x$
and vary $Q^2$ than when we fix $Q^2$ and vary $x$.

The next six figures repeat the previous distributions but
for $F_L(x,Q^2,m_c^2)$ rather than $F_2(x,Q^2,m_c^2)$. The general
features are much the same, but the integrated results are smaller for
$F_L(x,Q^2,m_c^2)$ than for $F_2(x,Q^2,m_c^2)$.
Figures 11(a) and 11(b) show the
distributions in the transverse momentum of the pair.
Figures 12(a) and 12(b) show the distributions in the
azimuthal angle between the ${\bf P}_t$ vectors of the
outgoing charm quark and charm antiquark. Then
we present in figs. 13(a) and 13(b) the histograms of the
distributions in $R$.

For completeness we now repeat the last twelve distributions
taking the heavy quark to be the bottom quark with mass $m_b= 4.75$
GeV$/c$ and the heavy-antiquark to be the bottom antiquark
with the same mass. The renormalization (factorization )
scale is chosen to be $\mu^2 = Q^2 + m_b^2 + (P_t^b+P_t^{\bar{b}})^2/4$.
Aside from the $P_t$ dependence, this choice reduces to the
usual choice of $\mu^2=Q^2$ for electroproduction of massless quarks and
$\mu^2=m_b^2$ for the photoproduction of bottom quarks.
Here we used the CTEQ3M ( $\overline{{\rm MS}}$ ) distributions and
the two loop running coupling with $\Lambda_5 = 0.158 {\rm GeV}$.


The integrated results are smaller for bottom quarks than for charm quarks
reflecting a decrease in $F_2$ by a factor of approximately
$50$, and in $F_L$ by a factor of approximately $150$.
We refer the interested reader to \cite{HS} for tables containing
$F_2(x,Q^2,m^2)$ and $F_L(x,Q^2,m^2)$ for both charm and bottom
quarks in the same $x$ and $Q^2$ bins presented here.
Apart from this decrease there are general features which are to
be expected.  In figs.\ 14(a), 14(b), 17(a), and 17(b)
we see that the $P_t$ spectra are harder.  The $\Delta \phi$ histograms
in figs.\ 15(a), 15(b), 18(a), and 18(b) show dominant back-to-back peaking,
which is stronger for bottom quarks than for charm quarks as the NLO
corrections are correspondingly smaller.
Finally in figs.\ 16(a), 16(b), 19(a), and 19(b) we see that the $R$
histograms are very $Q^2$ dependent.

In addition to checking that all plots presented are indeed independent of
the parameters $\tilde{\rho}$ and $\omega$ we have studied the renormalization
(factorization) scale dependence of the $P_t$, $\Delta \phi$,
and $R$ distributions presented here and the $M$ distributions
presented in \cite{HS}.  To aid the discussion we present tables 1-4
containing the averages of $P_t$, $\Delta \phi$, $R$, and $M$ as functions
of the renormalization (factorization) scale for charm production.
The subscripts $2$ and $L$ refer to averages over $F_2$ and $F_L$
respectively.  As customary we have presented averages for the
renormalization (factorization) scale $\mu=\mu_0$ chosen
in the previous plots and for $\mu=\mu_0/2$ and $\mu=2\mu_0$.
In tables 5-8 we show the same quantities for bottom production.
Typical scale variations from the central value are around
5 percent for charm production and 1 percent for bottom production.
In general the shapes of the $P_t$, $\Delta \phi$, and $R$
distributions remain fixed while we vary the scale but the normalization
changes.  This is to be expected because the corresponding plots are delta
functions at lowest order.  However, the $M$ distributions have
contributions from LO already so while the normalization stays roughly fixed
the shape of the plot changes slightly reflecting the change in the average
values in the tables.

Our programs can also produce two dimensional plots.  For example one
might study $dF_k(x,Q^2,m^2,y^Q,y^{\bar{Q}})/dy^Qdy^{\bar{Q}}$ or
$dF_k(x,Q^2,m^2,y^Q,M)/dy^QdM$ or other combinations.
In addition, experimental cuts can be implemented.

\mysection{Conclusion}

In this article we have outlined the NLO calculation of the
virtual-photon-parton (Wilson) coefficient functions in the
exclusive production of heavy quarks plus one jet.
This completes the study of the NLO electroproduction of heavy quarks.
The single particle inclusive calculation has already been published
\cite{LRSvN1}, \cite{LRSvN2}.  Also there are fits available in \cite{RSvN}
for the $\eta$ and $\xi$ dependence of the scale independent
coefficient functions allowing for a fast
numerical estimation of integrated rates for experiments.
In this series of papers we have concentrated
on deep-inelastic electroproduction where there is no need to introduce
any partonic densities in the photon.

The computer program we have written for the exclusive calculation
has the advantage that the four vectors of the heavy quark, heavy-antiquark
and/or one additional light parton jet are produced for each event
and can be subjected to experimental cuts.  We have not done this in any
of the plots shown here but the computer program is available and can be
easily modified to incorporate acceptances of the detectors at HERA.
\footnote{Requests for the computer
program should be sent to smith@elsebeth.physics.sunysb.edu.}
We have previously presented NLO distributions in the invariant masses
of the heavy quark heavy-antiquark pair \cite{HS} and shown that they
have reasonably smooth $K$-factors, i.e., one can generate
these distributions by multiplying the Born differential
cross sections by constant factors, usually around
$K= 1.3$ for bottom and $1.7$ for charm.  However, the study \cite{LRSvN2}
of the single particle inclusive distributions in $p_t$ and $y$ showed that
they do not have smooth $K$ factors.

For this paper we have presented those plots which depend on
information from the four vector of the additional jet.
We showed the distributions in the transverse-momentum
($P_t$) of the heavy quark antiquark pair, in the azimuthal angle
($\Delta \phi$) between the
${\bf P}_t$ vectors of the heavy quark and heavy antiquark, and in the
distribution in the heavy quark cone size ($R$).
All quantities were predicted in the c.\ m.\ frame of the photon-proton
system after integration over the azimuthal angle between the plane
containing the incoming and outgoing lepton and the plane containing the
incoming proton and outgoing heavy quark.  The results were presented as
distributions in $F_2(x,Q^2,m^2)$ and $F_L(x,Q^2,m^2)$
at specific points in $x$, $Q^2$ and
$m^2=m_c^2$ or $m^2 = m_b^2$.
None of these distributions can be reproduced
by any $K$-factor multiplication as the corresponding Born
distributions are proportional to delta-functions.
The spread of the tails in these
distributions indirectly measures the relative size of the NLO
contribution to that of the LO contribution.
In all cases the histograms of these
distributions have negative bins. These are regions where the
NLO calculation is not sufficient and a NNLO order
calculation (or some form of resummation) should be made.
A general statement about the magnitude of the NLO contribution
compared to the LO one is difficult to make as the size and sign of
the corrections may vary strongly between different regions of phase space.
However, we see that all plots have larger $Q^2$ variation at fixed $x$
as comparied to varying $x$ at fixed $Q^2$.
By varying the renormalization (factorization) scale we observed that
the distributions presented here changed in normalization but not in shape
while the invariant mass distribution presented earlier \cite{HS}
kept approximately the same normalization and had a mild shape change.
In our study we also calculated averages of various quantities and
found the typical variation between central and extreme scale choices
of 5 percent for charm production and 1 percent for bottom production.

\vskip 0.5 cm
\centerline{\bf Acknowledgments}
We acknowledge S. Mendoza for help in the early stages of the project.
We also thank E. Laenen, S. Riemersma, W. L. van Neerven, and
J. Whitmore  for helpful discussions and Michael Fischer for maintaining
our computers.
Our research is supported in part by the contract NSF 9309888.

\newpage

\newcommand{\tema}{$\times 10^{-1}$}
\newcommand{\temb}{$\times 10^{-2}$}
\newcommand{\temc}{$\times 10^{-3}$}
\newcommand{\temd}{$\times 10^{-4}$}
\newcommand{\teme}{$\times 10^{-5}$}
\newcommand{\temf}{$\times 10^{-5}$}

\centerline{\bf \large{Table 1}}
\vspace{2cm}

\begin{center}
\begin{tabular}{||c|c||c|c|c||c|c|c||}
\hline \hline
\multicolumn{2}{||c||}{Range}                        &
\multicolumn{3}{c||}{$ \langle P_t \rangle_2 $} &
\multicolumn{3}{c||}{$ \langle \Delta \phi \rangle_2 $}
\\ \hline \hline
$x$ & $Q^2$ & $\mu=\mu_0/2$ & $\mu=\mu_0$ & $\mu=2\mu_0$
            & $\mu=\mu_0/2$ & $\mu=\mu_0$ & $\mu=2\mu_0$ \\
\hline \hline
8.5 \temd & 8.5 & 2.73 & 1.82 & 1.35 & 1.45 & 2.01 & 2.32 \\ \hline
8.5 \temd & 12  & 3.02 & 2.04 & 1.51 & 1.42 & 1.97 & 2.28 \\ \hline
8.5 \temd & 25  & 3.72 & 2.59 & 1.95 & 1.39 & 1.91 & 2.22 \\ \hline
8.5 \temd & 50  & 4.53 & 3.24 & 2.48 & 1.40 & 1.88 & 2.17 \\ \hline
          &     &      &      &      &      &      &      \\ \hline
4.2 \temd & 12  & 3.63 & 2.45 & 1.80 & 1.16 & 1.79 & 2.15 \\ \hline
8.5 \temd & 12  & 3.02 & 2.04 & 1.51 & 1.42 & 1.97 & 2.28 \\ \hline
1.6 \temc & 12  & 2.52 & 1.71 & 1.28 & 1.63 & 2.12 & 2.39 \\ \hline
2.7 \temc & 12  & 2.14 & 1.46 & 1.10 & 1.79 & 2.23 & 2.47 \\ \hline
\hline
\end{tabular}
\end{center}
\vspace{2cm}

\begin{description}
\item[Table 1.]
Variation of
$ \langle P_t \rangle_2 $
and
$ \langle \Delta \phi \rangle_2 $
for charm production with
$\mu_0^2=Q^2+4(m_c^2+( P_t^c + P_t^{\bar{c}} )^2/4)$
for various $x$ and $Q^2$ values.
\end{description}
\newpage

\centerline{\bf \large{Table 2}}
\vspace{2cm}

\begin{center}
\begin{tabular}{||c|c||c|c|c||c|c|c||}
\hline \hline
\multicolumn{2}{||c||}{Range}                        &
\multicolumn{3}{c||}{$ \langle R \rangle_2 $} &
\multicolumn{3}{c||}{$ \langle M \rangle_2 $}
\\ \hline \hline
$x$ & $Q^2$ & $\mu=\mu_0/2$ & $\mu=\mu_0$ & $\mu=2\mu_0$
            & $\mu=\mu_0/2$ & $\mu=\mu_0$ & $\mu=2\mu_0$ \\
\hline \hline
8.5 \temd & 8.5 & 2.25 & 2.70 & 2.93 & 7.96 & 7.58 & 7.31 \\ \hline
8.5 \temd & 12  & 2.27 & 2.70 & 2.94 & 8.54 & 8.14 & 7.87 \\ \hline
8.5 \temd & 25  & 2.34 & 2.76 & 3.00 & 10.2 & 9.77 & 9.45 \\ \hline
8.5 \temd & 50  & 2.49 & 2.87 & 3.11 & 12.3 & 12.0 & 11.7 \\ \hline
          &     &      &      &      &      &      &      \\ \hline
4.2 \temd & 12  & 2.05 & 2.55 & 2.83 & 8.98 & 8.54 & 8.21 \\ \hline
8.5 \temd & 12  & 2.27 & 2.70 & 2.94 & 8.54 & 8.14 & 7.87 \\ \hline
1.6 \temc & 12  & 2.45 & 2.83 & 3.03 & 8.09 & 7.75 & 7.51 \\ \hline
2.7 \temc & 12  & 2.59 & 2.92 & 3.10 & 7.72 & 7.41 & 7.21 \\ \hline
\hline
\end{tabular}
\end{center}
\vspace{2cm}

\begin{description}
\item[Table 2.]
Variation of
$ \langle R \rangle_2 $
and
$ \langle M \rangle_2 $
for charm production with
$\mu_0^2=Q^2+4(m_c^2+( P_t^c + P_t^{\bar{c}} )^2/4)$
for various $x$ and $Q^2$ values.
\end{description}
\newpage

\centerline{\bf \large{Table 3}}
\vspace{2cm}

\begin{center}
\begin{tabular}{||c|c||c|c|c||c|c|c||}
\hline \hline
\multicolumn{2}{||c||}{Range}                        &
\multicolumn{3}{c||}{$ \langle P_t \rangle_L $} &
\multicolumn{3}{c||}{$ \langle \Delta \phi \rangle_L $}
\\ \hline \hline
$x$ & $Q^2$ & $\mu=\mu_0/2$ & $\mu=\mu_0$ & $\mu=2\mu_0$
            & $\mu=\mu_0/2$ & $\mu=\mu_0$ & $\mu=2\mu_0$ \\
\hline \hline
8.5 \temd & 8.5 & 2.47 & 1.73 & 1.31 & 2.17 & 2.45 & 2.62 \\ \hline
8.5 \temd & 12  & 2.78 & 1.96 & 1.50 & 2.14 & 2.42 & 2.59 \\ \hline
8.5 \temd & 25  & 3.58 & 2.60 & 2.00 & 2.12 & 2.38 & 2.55 \\ \hline
8.5 \temd & 50  & 4.60 & 3.44 & 2.68 & 2.13 & 2.36 & 2.53 \\ \hline
          &     &      &      &      &      &      &      \\ \hline
4.2 \temd & 12  & 3.25 & 2.30 & 1.75 & 2.02 & 2.33 & 2.52 \\ \hline
8.5 \temd & 12  & 2.78 & 1.96 & 1.50 & 2.14 & 2.42 & 2.59 \\ \hline
1.6 \temc & 12  & 2.37 & 1.68 & 1.28 & 2.25 & 2.51 & 2.66 \\ \hline
2.7 \temc & 12  & 2.05 & 1.45 & 1.12 & 2.34 & 2.57 & 2.71 \\ \hline
\hline
\end{tabular}
\end{center}
\vspace{2cm}

\begin{description}
\item[Table 3.]
Variation of
$ \langle P_t \rangle_L $
and
$ \langle \Delta \phi \rangle_L $
for charm production with
$\mu_0^2=Q^2+4(m_c^2+( P_t^c + P_t^{\bar{c}} )^2/4)$
for various $x$ and $Q^2$ values.
\end{description}
\newpage

\centerline{\bf \large{Table 4}}
\vspace{2cm}

\begin{center}
\begin{tabular}{||c|c||c|c|c||c|c|c||}
\hline \hline
\multicolumn{2}{||c||}{Range}                        &
\multicolumn{3}{c||}{$ \langle R \rangle_L $} &
\multicolumn{3}{c||}{$ \langle M \rangle_L $}
\\ \hline \hline
$x$ & $Q^2$ & $\mu=\mu_0/2$ & $\mu=\mu_0$ & $\mu=2\mu_0$
            & $\mu=\mu_0/2$ & $\mu=\mu_0$ & $\mu=2\mu_0$ \\
\hline \hline
8.5 \temd & 8.5 & 2.59 & 2.78 & 2.90 & 6.15 & 5.97 & 5.86 \\ \hline
8.5 \temd & 12  & 2.58 & 2.77 & 2.89 & 6.56 & 6.36 & 6.23 \\ \hline
8.5 \temd & 25  & 2.58 & 2.77 & 2.89 & 7.75 & 7.53 & 7.39 \\ \hline
8.5 \temd & 50  & 2.62 & 2.79 & 2.91 & 9.42 & 9.17 & 9.04 \\ \hline
          &     &      &      &      &      &      &      \\ \hline
4.2 \temd & 12  & 2.01 & 2.70 & 2.84 & 6.80 & 6.51 & 6.34 \\ \hline
8.5 \temd & 12  & 2.58 & 2.77 & 2.89 & 6.56 & 6.63 & 6.23 \\ \hline
1.6 \temc & 12  & 2.65 & 2.83 & 2.94 & 6.39 & 6.21 & 6.11 \\ \hline
2.7 \temc & 12  & 2.71 & 2.87 & 2.97 & 6.21 & 6.09 & 5.99 \\ \hline
\hline
\end{tabular}
\end{center}
\vspace{2cm}

\begin{description}
\item[Table 4.]
Variation of
$ \langle R \rangle_L $
and
$ \langle M \rangle_L $
for charm production with
$\mu_0^2=Q^2+4(m_c^2+( P_t^c + P_t^{\bar{c}} )^2/4)$
for various $x$ and $Q^2$ values.
\end{description}
\newpage

\centerline{\bf \large{Table 5}}
\vspace{2cm}

\begin{center}
\begin{tabular}{||c|c||c|c|c||c|c|c||}
\hline \hline
\multicolumn{2}{||c||}{Range}                        &
\multicolumn{3}{c||}{$ \langle P_t \rangle_2 $} &
\multicolumn{3}{c||}{$ \langle \Delta \phi \rangle_2 $}
\\ \hline \hline
$x$ & $Q^2$ & $\mu=\mu_0/2$ & $\mu=\mu_0$ & $\mu=2\mu_0$
            & $\mu=\mu_0/2$ & $\mu=\mu_0$ & $\mu=2\mu_0$ \\
\hline \hline
8.5 \temd & 8.5 & 3.52 & 2.54 & 1.94 & 2.29 & 2.52 & 2.66 \\ \hline
8.5 \temd & 12  & 3.88 & 2.83 & 2.17 & 2.24 & 2.48 & 2.63 \\ \hline
8.5 \temd & 25  & 4.64 & 3.47 & 2.70 & 2.17 & 2.41 & 2.56 \\ \hline
8.5 \temd & 50  & 5.36 & 4.11 & 3.25 & 2.13 & 2.35 & 2.51 \\ \hline
          &     &      &      &      &      &      &      \\ \hline
4.2 \temd & 12  & 4.70 & 3.48 & 2.68 & 2.12 & 2.37 & 2.54 \\ \hline
8.5 \temd & 12  & 3.88 & 2.83 & 2.17 & 2.24 & 2.48 & 2.63 \\ \hline
1.6 \temc & 12  & 3.16 & 2.27 & 1.74 & 2.35 & 2.57 & 2.70 \\ \hline
2.7 \temc & 12  & 2.57 & 1.82 & 1.40 & 2.45 & 2.65 & 2.76 \\ \hline
\hline
\end{tabular}
\end{center}
\vspace{2cm}

\begin{description}
\item[Table 5.]
Variation of
$ \langle P_t \rangle_2 $
and
$ \langle \Delta \phi \rangle_2 $
for bottom production with
$\mu_0^2=Q^2+m_b^2+( P_t^b + P_t^{\bar{b}} )^2/4$
for various $x$ and $Q^2$ values.
\end{description}
\newpage

\centerline{\bf \large{Table 6}}
\vspace{2cm}

\begin{center}
\begin{tabular}{||c|c||c|c|c||c|c|c||}
\hline \hline
\multicolumn{2}{||c||}{Range}                        &
\multicolumn{3}{c||}{$ \langle R \rangle_2 $} &
\multicolumn{3}{c||}{$ \langle M \rangle_2 $}
\\ \hline \hline
$x$ & $Q^2$ & $\mu=\mu_0/2$ & $\mu=\mu_0$ & $\mu=2\mu_0$
            & $\mu=\mu_0/2$ & $\mu=\mu_0$ & $\mu=2\mu_0$ \\
\hline \hline
8.5 \temd & 8.5 & 2.77 & 2.96 & 3.07 & 16.7 & 16.6 & 16.4 \\ \hline
8.5 \temd & 12  & 2.74 & 2.93 & 3.05 & 17.3 & 17.2 & 17.0 \\ \hline
8.5 \temd & 25  & 2.71 & 2.89 & 3.02 & 18.8 & 18.6 & 18.4 \\ \hline
8.5 \temd & 50  & 2.71 & 2.88 & 3.01 & 20.5 & 20.3 & 20.1 \\ \hline
          &     &      &      &      &      &      &      \\ \hline
4.2 \temd & 12  & 2.64 & 2.85 & 2.98 & 18.3 & 18.1 & 17.8 \\ \hline
8.5 \temd & 12  & 2.74 & 2.93 & 3.05 & 17.3 & 17.2 & 17.0 \\ \hline
1.6 \temc & 12  & 2.83 & 3.00 & 3.11 & 16.4 & 16.3 & 16.1 \\ \hline
2.7 \temc & 12  & 2.90 & 3.06 & 3.15 & 15.6 & 15.5 & 15.4 \\ \hline
\hline
\end{tabular}
\end{center}
\vspace{2cm}

\begin{description}
\item[Table 6.]
Variation of
$ \langle R \rangle_2 $
and
$ \langle M \rangle_2 $
for bottom production with
$\mu_0^2=Q^2+m_b^2+( P_t^b + P_t^{\bar{b}} )^2/4$
for various $x$ and $Q^2$ values.
\end{description}
\newpage

\centerline{\bf \large{Table 7}}
\vspace{2cm}

\begin{center}
\begin{tabular}{||c|c||c|c|c||c|c|c||}
\hline \hline
\multicolumn{2}{||c||}{Range}                        &
\multicolumn{3}{c||}{$ \langle P_t \rangle_L $} &
\multicolumn{3}{c||}{$ \langle \Delta \phi \rangle_L $}
\\ \hline \hline
$x$ & $Q^2$ & $\mu=\mu_0/2$ & $\mu=\mu_0$ & $\mu=2\mu_0$
            & $\mu=\mu_0/2$ & $\mu=\mu_0$ & $\mu=2\mu_0$ \\
\hline \hline
8.5 \temd & 8.5 & 2.89 & 2.24 & 1.80 & 2.65 & 2.75 & 2.82 \\ \hline
8.5 \temd & 12  & 3.29 & 2.54 & 2.03 & 2.62 & 2.72 & 2.80 \\ \hline
8.5 \temd & 25  & 4.08 & 3.18 & 2.55 & 2.56 & 2.68 & 2.76 \\ \hline
8.5 \temd & 50  & 4.86 & 3.85 & 3.13 & 2.52 & 2.64 & 2.73 \\ \hline
          &     &      &      &      &      &      &      \\ \hline
4.2 \temd & 12  & 3.91 & 3.05 & 2.46 & 2.55 & 2.67 & 2.75 \\ \hline
8.5 \temd & 12  & 3.29 & 2.54 & 2.03 & 2.62 & 2.72 & 2.80 \\ \hline
1.6 \temc & 12  & 2.72 & 2.08 & 1.67 & 2.67 & 2.77 & 2.84 \\ \hline
2.7 \temc & 12  & 2.25 & 1.72 & 1.38 & 2.72 & 2.81 & 2.87 \\ \hline
\hline
\end{tabular}
\end{center}
\vspace{2cm}

\begin{description}
\item[Table 7.]
Variation of
$ \langle P_t \rangle_L $
and
$ \langle \Delta \phi \rangle_L $
for bottom production with
$\mu_0^2=Q^2+m_b^2+( P_t^b + P_t^{\bar{b}} )^2/4$
for various $x$ and $Q^2$ values.
\end{description}
\newpage

\centerline{\bf \large{Table 8}}
\vspace{2cm}

\begin{center}
\begin{tabular}{||c|c||c|c|c||c|c|c||}
\hline \hline
\multicolumn{2}{||c||}{Range}                        &
\multicolumn{3}{c||}{$ \langle R \rangle_L $} &
\multicolumn{3}{c||}{$ \langle M \rangle_L $}
\\ \hline \hline
$x$ & $Q^2$ & $\mu=\mu_0/2$ & $\mu=\mu_0$ & $\mu=2\mu_0$
            & $\mu=\mu_0/2$ & $\mu=\mu_0$ & $\mu=2\mu_0$ \\
\hline \hline
8.5 \temd & 8.5 & 2.91 & 2.98 & 3.02 & 15.0 & 14.9 & 14.8 \\ \hline
8.5 \temd & 12  & 2.88 & 2.95 & 3.00 & 15.3 & 15.1 & 15.0 \\ \hline
8.5 \temd & 25  & 2.84 & 2.92 & 2.98 & 16.0 & 15.9 & 15.8 \\ \hline
8.5 \temd & 50  & 2.82 & 2.90 & 2.96 & 17.1 & 17.0 & 16.8 \\ \hline
          &     &      &      &      &      &      &      \\ \hline
4.2 \temd & 12  & 2.84 & 2.91 & 2.97 & 15.7 & 15.5 & 15.3 \\ \hline
8.5 \temd & 12  & 2.88 & 2.95 & 3.00 & 15.3 & 15.1 & 15.0 \\ \hline
1.6 \temc & 12  & 2.92 & 2.99 & 3.04 & 14.8 & 14.8 & 14.7 \\ \hline
2.7 \temc & 12  & 2.96 & 3.02 & 3.06 & 14.4 & 14.4 & 14.3 \\ \hline
\hline
\end{tabular}
\end{center}
\vspace{2cm}

\begin{description}
\item[Table 8.]
Variation of
$ \langle R \rangle_L $
and
$ \langle M \rangle_L $
for bottom production with
$\mu_0^2=Q^2+m_b^2+( P_t^b + P_t^{\bar{b}} )^2/4$
for various $x$ and $Q^2$ values.
\end{description}
\newpage

\appendix

%
%
\newpage
\mysection*{Appendix A}
\setcounter{section}{1}

Here we discuss the kinematic variables and phase space
in the $2 \rightarrow 2$ and $2 \rightarrow 3$ reactions.
We then define the generalized plus distributions used in the text.
The Mandelstam invariants for the reaction,
\begin{eqnarray}
\gamma^{\ast} (q) + a_1(k_1) \rightarrow Q(p_1) + \overline{Q}(p_2),
\end{eqnarray}
with $k_1^2=0$ and $p_i^2=m^2$ are
\begin{eqnarray}
s^{\prime}  & \equiv & s-q^2 = (q+k_1)^2-q^2 = 2q\cdot k_1, \nonumber \\
t_1         & \equiv & t-m^2 = (k_1-p_2)^2-m^2 = -2k_1\cdot p_2, \nonumber \\
u_1         & \equiv & u-m^2 = (q-p_2)^2-m^2 = -2q\cdot p_2 + q^2,
\end{eqnarray}
which satisfy $s^{\prime}+t_1+u_1=0$.
The two body phase space in space-time dimension $n=4+\epsilon$ is
\begin{equation}
\label{sp2}
d \Gamma_2 = \frac{2^{-\epsilon}}{16 \pi} \left( \frac{s}{4 \pi} \right)
^{\epsilon/2} \beta^{1+\epsilon} \frac{1}{\Gamma(1+\epsilon/2)}
\sin^{1+\epsilon} \theta_1 d \theta_1,
\end{equation}
where $\beta = \sqrt{1-\rho}, \rho = 4m^2/s$ and $\theta_1$ is the
angle between
${\bf q}$ and ${\bf p}_1$ in the $\gamma^{\ast} a_1$ center-of-mass frame.
Therefore we have
\begin{eqnarray}
t_1 & = & - \frac{1}{2} s^{\prime} (1-\beta \cos \theta_1 ), \nonumber \\
u_1 & = & - \frac{1}{2} s^{\prime} (1+\beta \cos \theta_1 ),
\end{eqnarray}
with $0 \leq \theta_1 \leq \pi$.

For the two to three body process
\begin{eqnarray}
\gamma^{\ast} (q) + a_1(k_1) \rightarrow Q(p_1) + \overline{Q}(p_2) + a_2(k_2),
\end{eqnarray}
with $k_i^2=0$ and $p_i^2=m^2$ there are 5 independent invariants which
we take to be
\begin{eqnarray}
s^{\prime}   & = & s - q^2 = (q+k_1)^2-q^2 = 2q\cdot k_1, \nonumber \\
t_1          & = & (k_1-p_2)^2 - m^2 = -2k_1\cdot p_2, \nonumber \\
	u^{\prime}_1 & \equiv & u_1 - q^2 = (q-p_2)^2-m^2-q^2 = -2q\cdot p_2,
	\nonumber \\
t^{\prime}   & = & (k_1-k_2)^2 = -2k_1\cdot k_2, \nonumber \\
u^{\prime}   & = & (q-k_2)^2 = q^2 - 2 q\cdot k_2,
\end{eqnarray}
where $q+k_1=k_2+p_1+p_2$.
The remaining five invariants are
\begin{eqnarray}
s_3 & = & (k_2+p_2)^2-m^2 =-s^{\prime}-u^{\prime}-t^{\prime}-u^{\prime}_1-t_1
= 2k_2\cdot p_2,
\nonumber \\
s_4 & = & (k_2+p_1)^2-m^2 = s^{\prime}+u_1+t_1 = 2k_2\cdot p_1, \nonumber \\
s_5 & = & (p_1+p_2)^2 = s^{\prime}+u^{\prime}+t^{\prime}
=  2m^2 + 2p_1\cdot p_2, \nonumber \\
u_6 & = & (k_1-p_1)^2-m^2 = -s^{\prime}-t^{\prime}-t_1 = - 2k_1\cdot p_1,
\nonumber \\
u_7 & = & (q-p_1)^2-m^2 = -s^{\prime}-u^{\prime}-u^{\prime}_1
= q^2 - 2p_1\cdot q \, .
\end{eqnarray}
We introduce the variables $x=s^{\prime}_5/s^{\prime}$ where
$s^{\prime}_5=s_5-q^2$, and $y$ is the cosine of the angle
between ${\bf q}$ and ${\bf k}_2$ in the $\gamma^{\ast} a_1$
center-of-mass frame.  These have ranges
$\rho^{\ast} \leq x \leq 1$ and $-1 \leq y \leq 1$
with $\rho^{\ast} = (4m^2-q^2)/s^{\prime}$.  We then find
\begin{eqnarray}
t^{\prime} & = & -\frac{1}{2} s^{\prime} \left( \frac{s^{\prime}}{s} \right)
(1-x) (1+y), \nonumber \\
u^{\prime} & = & q^2 \left( \frac{s_5}{s} \right) - \frac{1}{2} s^{\prime}
\left( \frac{s^{\prime}}{s} \right) (1-x) (1-y).
\end{eqnarray}
In the center-of-mass system of the outgoing heavy quark antiquark pair
we decompose the momenta as follows:
\begin{eqnarray}
q   & = & (q^{0},0,0,\mid \! {\bf q} \! \mid ), \nonumber \\
k_1 & = & k_1^0(1,0, \sin \psi, \cos \psi), \nonumber \\
k_2 & = & (k_2^0,0,k_1^0 \sin \psi, \mid \! {\bf q} \! \mid + k_1^0
\cos \psi ), \nonumber \\
p_1 & = & \frac{1}{2} \sqrt{s_5} (1,\beta_5 \sin \theta_2 \sin \theta_1,
\beta_5 \cos \theta_2 \sin \theta_1, \beta_5 \cos \theta_1 ), \nonumber \\
p_2 & = & \frac{1}{2} \sqrt{s_5} (1,-\beta_5 \sin \theta_2 \sin \theta_1,
-\beta_5 \cos \theta_2 \sin \theta_1, -\beta_5 \cos \theta_1 ),
\end{eqnarray}
where
\begin{eqnarray}
q^0 & = & \frac{s+u^{\prime}}{2 \sqrt{s_5}}, \nonumber \\
\mid \! {\bf q} \! \mid & = & \frac{1}{2 \sqrt{s_5}}
                    \sqrt{(s+u^{\prime})^2-4s_5q^2}, \nonumber \\
k_1^0 & = & \frac{s_5-u^{\prime}}{2 \sqrt{s_5}}, \nonumber \\
k_2^0 & = & \frac{s-s_5}{2 \sqrt{s_5}}, \nonumber \\
\cos \psi & = & \frac{-s^{\prime}+2 k_1^0 q^0}{2 k_1^0 \mid
                   \! {\bf q} \! \mid}, \nonumber \\
\beta_5 & = & \sqrt{1-4m^2/s_5} .
\end{eqnarray}
Note that in the limit $x \rightarrow 1$, $\beta_5 \rightarrow \beta$.
The remaining two independent invariants are
\begin{eqnarray}
t_1 & = & -\frac{1}{2}(s_5-u^{\prime})(1
        +\beta_5 \cos \theta_2 \sin \theta_1 \sin \psi
        +\beta_5 \cos \theta_1 \cos \psi ), \nonumber \\
u_1 & = & q^2 -\frac{1}{2}(s+u^{\prime}+2 \sqrt{s_5}
        \beta_5 \mid \! {\bf q} \! \mid \cos \theta_1 ).
\end{eqnarray}
We will also need the following expressions
\begin{eqnarray*}
s_3 \! & = & \! \frac{1}{2} \left[ s-s_5+(s_5-u^{\prime})
          \beta_5 \cos \theta_2 \sin \theta_1 \sin \psi +
          \beta_5 \cos \theta_1 ( 2 \sqrt{s_5} \mid \! {\bf q} \! \mid +
          (s_5-u^{\prime}) \cos \psi ) \right], \nonumber \\
s_4 \! & = & \! \frac{1}{2} \left[ s-s_5-(s_5-u^{\prime})
          \beta_5 \cos \theta_2 \sin \theta_1 \sin \psi -
          \beta_5 \cos \theta_1 ( 2 \sqrt{s_5} \mid \! {\bf q} \! \mid +
          (s_5-u^{\prime}) \cos \psi ) \right].
\end{eqnarray*}

The three body phase space in $n=4+\epsilon$ space-time dimensions
expressed in terms of $x,y,\theta_1$, and $\theta_2$ is
\begin{equation}
\label{ps3}
d\Gamma_3 = HNd\Gamma_2^{(5)} \frac{(s^{\prime})^{1+\epsilon/2}}{2 \pi}
\left( \frac{s^{\prime}}{s} \right)^{1+\epsilon/2} (1-x)^{1+\epsilon}
(1-y^2)^{\epsilon/2} dy \sin^{\epsilon} \theta_2 d \theta_2,
\end{equation}
with $0 \leq \theta_1 \leq \pi$,
$0 \leq \theta_2 \leq \pi$, $\rho^{\ast} \leq x \leq 1$,
$-1 \leq y \leq 1$,  and
\begin{eqnarray}
H & = &  \frac{\Gamma(1+\epsilon/2)}{\Gamma(1-\epsilon/2)\Gamma(1+\epsilon)}
	= 1 - \frac{\pi^2}{12} \epsilon^2 + O( \epsilon^3),  \\
N & = & \frac{(4 \pi)^{-\epsilon/2}}{(4 \pi)^2} \Gamma(1-\epsilon/2)
	= \frac{1}{16 \pi^2} \left( \frac{\epsilon}{2} \right) \left(
	\frac{2}{\overline{\epsilon}} \right) + O(\epsilon^2),
\end{eqnarray}
\begin{equation}
\label{ps25}
d\Gamma_2^{(5)} = \frac{2^{-\epsilon}}{16 \pi} \left( \frac{s_5}{4 \pi}
	\right)^{\epsilon/2} \beta_5^{1+\epsilon} \frac{1}{\Gamma( 1 +
	\epsilon/2)} \sin^{1+\epsilon} \theta_1 d \theta_1 dx,
\end{equation}
where $2/ \overline{\epsilon} = 2/ \epsilon + \gamma_E - \ln 4 \pi$.

The generalized plus distributions encountered in sec.\ 3 arise by making the
replacements
\begin{equation}
\label{yplus1}
(1+y)^{-1+  \epsilon} \sim
	\left( \frac{1}{1+y} \right)_{\omega} + \delta (1+y)
	\left( \frac{1}{\epsilon} + \ln \omega \right)
	+ O(\epsilon),
\end{equation}
\begin{eqnarray}
\label{yplus2}
(1-y^2)^{-1+  \epsilon} & \sim &
	\frac{1}{2} \left[ \left( \frac{1}{1+y} \right)_{\omega}
	+ \left( \frac{1}{1-y} \right)_{\omega} \right] \nonumber \\
	&+& \left[ \delta (1+y) + \delta (1-y) \right]
	  \left( \frac{1}{2\epsilon} + \frac{1}{2} \ln 2 \omega \right)
	+ O(\epsilon),
\end{eqnarray}
\begin{eqnarray}
\label{xplus}
(1-x)^{-1+ \epsilon} & \sim &
	\left( \frac{1}{1-x} \right)_{\tilde{\rho}} +  \epsilon \left(
	\frac{\ln(1-x)}{1-x} \right)_{\tilde{\rho}} \nonumber \\
	&+& \delta (1-x) \left[ \frac{1}{ \epsilon} + 2 \ln \tilde{\beta}
	+ 2 \epsilon \ln^2 \tilde{\beta} \right]
	+ O(\epsilon^2) ,
\end{eqnarray}
inside integrations over smooth functions.  We have defined
$\tilde{\beta} = \sqrt{1-\tilde{\rho}}$, and the generalized plus
distributions are defined by
\begin{eqnarray}
\int^1_{\tilde{\rho}} dx f(x) \left( \frac{1}{1-x} \right)_{\tilde{\rho}}
	& = & \int^1_{\tilde{\rho}} dx  \frac{f(x)-f(1)}{1-x},  \\
\int^1_{\tilde{\rho}} dx f(x) \left( \frac{\ln(1-x)}{1-x}
	\right)_{\tilde{\rho}} & = &
	\int^1_{\tilde{\rho}} dx  \frac{f(x)-f(1)}{1-x} \ln(1-x), \\
\int^{-1+\omega}_{-1} dy f(y) \left( \frac{1}{1+y} \right)_{\omega} & = &
	\int^{-1+\omega}_{-1} dy \frac{f(y)-f(-1)}{1+y}, \\
\int_{1-\omega}^{1} dy f(y) \left( \frac{1}{1-y} \right)_{\omega} & = &
	\int_{1-\omega}^{1} dy \frac{f(y)-f(1)}{1-y} \, ,
\end{eqnarray}
where $\rho^{\ast} \leq \tilde{\rho} < 1$ and $0 < \omega \leq 2$.
Thus we see, in distinction from the phase space slicing method, that the soft
and collinear parameters $\tilde{\rho}$ and $\omega$
are {\em not} required to be small.
It is understood that when the integration range does not enclose a
singularity, the distribution sign is dropped (c.f. (\ref{distex})).

%
%
\newpage
\mysection*{Appendix B}
\setcounter{section}{2}

In this appendix we discuss the derivation of the soft limit $x=1$ of the
matrix element $M_i^g(3)$.  We use standard techniques for the emission of
soft gluons \cite{muta}, \cite{mnr}, \cite{ks}.

Consider the diagram shown in fig.\ 3(d).
The momenta are defined according to fig.\ 1.
Take the incoming gluon to have color index $a$ and Lorentz index $\mu$,
the outgoing gluon to have color index $c$ and Lorentz index $\sigma$,
and the photon to have Lorentz index $\rho$.
Using standard Feynman rules one finds
\begin{equation}
I = -igf^{abc} V_{\mu \sigma \alpha}(k_1,-k_2,k_2-k_1)
\frac{\epsilon^{\mu}(k_1)}{2 k_1 \cdot k_2} B^{b \alpha \rho}
(q,k_1-k_2),
\end{equation}
where
\begin{equation}
V_{\mu \sigma \alpha}(k_1,-k_2,k_2-k_1) = (k_1+k_2)_{\alpha} g_{\mu \sigma}
+ (k_1-2k_2)_{\mu} g_{\sigma \alpha} + (k_2-2k_1)_{\sigma} g_{\mu \alpha} .
\end{equation}
$B_{ij}^{b \alpha \beta}$ is the sum of the lowest order diagrams in fig.\ 2
and we have chosen the Feynman gauge.  Now by using current conservation
$\left[ (k_1-k_2)_{\alpha} B^{b \alpha \rho} (q,k_1-k_2) = 0 \right]$,
and the transversality of the gluon polarization tensor
$\left[ \epsilon_{\mu}(k_1) k_1^{\mu} =0 \right]$,
and recalling that in the soft limit $k_2 \rightarrow 0$, a short
computation shows that
\begin{equation}
I = igf^{abc} \frac{k_1^{\sigma}}{k_1 \cdot k_2}
\epsilon_{\mu} (k_1) B^{b \mu \rho} (q,k_1)
+ {\rm finite \, terms \, as} \, k_2 \rightarrow 0.
\end{equation}
As we work in the Feynman gauge we must consider the interference
between all diagrams in fig.\ 3 \cite{hump}.  Analyzing them as we have
fig.\ 3(d) leads to
\begin{equation}
\label{Mgsoft}
M_i^g(3) = 8 g^4 e^2 e^2_H N C_F \left[ C_A S_{i,OK} + 2 C_F S_{i,QED} \right]
\end{equation}
in the soft limit when one neglects terms that are finite when multiplied
by the $(1-x)^{1+\epsilon}$ term in the three body phase space.  We have
defined
\begin{eqnarray}
\label{Sik}
S_{i,OK}  & = & \left[ (k_1p_2)+(p_1k_1)-(p_1p_2)     \right] B_{i,QED},
\nonumber \\
S_{i,QED} & = & \left[ (p_1p_2)-\frac{1}{2}(p_1p_1)-\frac{1}{2}(p_2p_2)
\right] B_{i,QED}.
\end{eqnarray}
and denoted $(uv)=u \cdot v / ( u \cdot k_2 \, v \cdot k_2 )$ which is
often called the eikonal factor.  Expanding the factors $(uv)$ we reproduce
eqs. (3.21) and (3.22) of \cite{LRSvN1}.  Recalling the definition of $
f_i^g(\theta_1)$ we find
\begin{eqnarray}
\label{ftheta1}
f_i^g(\theta_1) &=& \left[ \frac{1}{4} (s^{\prime})^2 \left(
\frac{s^{\prime}}{s} \right)^2 \right] \left[ 8 g^4 e^2 e^2_H N C_F \right]
B_{i,QED} \nonumber \\ & \times &
\left\{ C_A \left[ I_{(k_1p_2)} + I_{(p_1k_1)} - I_{(p_1p_2)} \right]
+ 2 C_F \left[ I_{(p_1p_2)} - \frac{1}{2} I_{(p_1p_1)}
- \frac{1}{2}I_{(p_2p_2)} \right] \right\}, \nonumber \\
&&
\end{eqnarray}
with
\begin{equation}
I_{(uv)} \equiv \int_{-1}^1 (1-y^2)^{-1+\epsilon/2} dy \int_0^{\pi}
\sin^{\epsilon} \theta_2 d \theta_2 \left. \left[ (uv) (1-x)^2 (1-y^2)
\right] \right|_{x=1}.
\end{equation}
Using the expansion of $s_3$ and $s_4$ about $x=1$
\begin{eqnarray}
s_3 & = & \frac{1}{2}s^{\prime}(1-x)\left[1+\beta \sqrt{1-y^2} \cos
        \theta_2 \sin \theta_1+\beta y \cos \theta_1 \right]+O( (1-x)^2 ),
\nonumber \\
s_4 & = & \frac{1}{2}s^{\prime}(1-x)\left[1-\beta \sqrt{1-y^2} \cos
        \theta_2 \sin \theta_1-\beta y \cos \theta_1 \right]+O( (1-x)^2 ),
\nonumber
\end{eqnarray}
we can write the integrals $I_{(uv)}$ in terms of
\begin{equation}
I^{(j)}_n \equiv \int_{0}^{\pi} d \alpha (\sin \alpha)^{n-3}
    \int_{0}^{\pi} d \beta (\sin \beta)^{n-4}
    (A+B\cos\alpha+C\sin\alpha\cos\beta)^{-j},
\end{equation}
with $j=1,2$ and $A^2 \neq B^2 + C^2$, and
\begin{equation}
\overline{I}^{(i,j)}_n \equiv \int_{0}^{\pi} d \alpha (\sin \alpha)^{n-3}
    \int_{0}^{\pi} d \beta (\sin \beta)^{n-4}
    (A+B\cos\alpha+C\sin\alpha\cos\beta)^{-j}
    (a+b \cos \alpha)^{-i},
\end{equation}
with $A^2 \neq B^2+C^2$, $a=-b$, and $i=j=1$.
The second integral may be found in the literature \cite{BKvNS} and the
first evaluated using the methods of \cite{willy}.  We find
\begin{eqnarray}
I^{(1)}_n & = & \frac{\pi}{\sqrt{B^2+C^2}}
\left\{ \ln \left( \frac{A+\sqrt{B^2+C^2}}{A-\sqrt{B^2+C^2}} \right)
\right. \nonumber \\ & & \left. -(n-4) \left[ {\rm Li_2} \left(
\frac{2\sqrt{B^2+C^2}}{A+\sqrt{B^2+C^2}}
\right)+\frac{1}{4} \ln^2 \left( \frac{A+\sqrt{B^2+C^2}}{A-\sqrt{B^2+C^2}}
\right) \right] \right\}, \nonumber \\
&&
\end{eqnarray}
and
\begin{eqnarray}
I^{(2)}_n & = & \frac{2 \pi}{A^2-B^2-C^2} \left[ 1 - \frac{1}{2} (n-4)
\frac{A}{\sqrt{B^2+C^2}} \ln \left(
\frac{A+\sqrt{B^2+C^2}}{A-\sqrt{B^2+C^2}} \right) \right], \nonumber \\
&&
\end{eqnarray}
where we drop $O((n-4)^2)$ terms and quote, for completeness, the result
for the second integral \cite{BKvNS}
\begin{eqnarray}
\overline{I}^{(1,1)}_n & = & \frac{\pi}{a(A+B)} \left\{ \frac{2}{n-4} +
\ln \left( \frac{(A+B)^2}{A^2-B^2-C^2} \right) \right. \nonumber \\
& + & \left.
\frac{1}{2} (n-4) \left[ \ln^2 \left( \frac{A-\sqrt{B^2+C^2}}{A+B} \right)
- \frac{1}{2} \ln^2 \left( \frac{A+\sqrt{B^2+C^2}}{A-\sqrt{B^2+C^2}} \right)
\right. \right. \nonumber \\ & + & \left. \left. 2 {\rm Li_2}
\left( - \frac{B+\sqrt{B^2+C^2}}{A-\sqrt{B^2+C^2}}
\right) - 2 {\rm Li_2} \left(  \frac{B-\sqrt{B^2+C^2}}{A+B} \right)
\right] \right\},
\end{eqnarray}
again dropping $O((n-4)^2)$ terms.  The dilogarithmic function ${\rm Li_2}(x)$
is defined in \cite{dilog}.

{}From these integrals and (\ref{Mgsoft}), (\ref{Sik}), (\ref{ftheta1}), and
(\ref{gs}) we find
\begin{equation}
d \sigma^{(s)}_{i,g} = C_{i,g} M_i^{soft} d \Gamma_2
\end{equation}
with
\begin{equation}
M_i^{soft} = 8 g^4 e^2 e^2_H N \mu^{-\epsilon} C_{\epsilon} C_F
	\left[ C_A \tilde{S}_{i,OK} + 2 C_F \tilde{S}_{i,QED} \right]
	B_{i,QED}.
\end{equation}
The factor
\begin{equation}
C_{\epsilon} = \frac{1}{16 \pi^2} e^{\epsilon (\gamma_E-\ln 4 \pi) / 2}
\left( \frac{ \mu^2 }{m^2} \right)^{ - \epsilon / 2}
\end{equation}
is common not only to $d \sigma_{i,g}^{(s)}$ but to $d \sigma_{i,g}^{(v)}$
as well (see Appendix
A of \cite{LRSvN1} ).  The mass parameter $\mu$ originates from the
dimensionality of the gauge coupling in $n$ dimensions.  The remaining terms
are
\begin{eqnarray}
\label{SOK}
\tilde{S}_{i,OK}
&=& \frac{8}{\epsilon^2} - \frac{4}{\epsilon} \left[
\frac{1}{\beta} \left( \frac{2m^2}{s}-1 \right) \ln \frac{1-\beta}{1+\beta}
-4 \ln \tilde{\beta} - \ln \frac{-t_1}{m^2} - \ln \frac{-u_1}{m^2} \right]
\nonumber \\
&-& \left[ \ln \frac{s}{m^2} + 2 \ln \frac{-t_1}{m^2}
+ 2 \ln \frac{-u_1}{m^2} -4 \ln \frac{s^{\prime}}{m^2} \right]
\ln \frac{s}{m^2}
\nonumber \\
&+& 4 \left[ \ln \frac{-t_1}{m^2}
+ \ln \frac{-u_1}{m^2} - \ln \frac{s^{\prime}}{m^2} \right]
 \ln \frac{s^{\prime}}{m^2}
- \frac{1}{\beta} \left( \frac{2m^2}{s}-1 \right)
\nonumber \\
& \times & \left\{
\left[ 8 \ln \tilde{\beta} -2 \ln \frac{s}{m^2}
+4 \ln \frac{s^{\prime}}{m^2} + \ln \frac{1-\beta}{1+\beta} \right]
  \ln \frac{1-\beta}{1+\beta} +
4 {\rm Li_2} \left( \frac{2 \beta}{1+\beta} \right) \right\}
\nonumber \\
&+& 8 \left( \ln \frac{-t_1}{m^2} + \ln \frac{-u_1}{m^2}
\right) \ln \tilde{\beta}
+ 16 \ln^2 \tilde{\beta} - 3 \zeta (2) - \ln^2
\frac{1-\beta}{1+\beta}
\nonumber \\
&+& 2 {\rm Li_2} \left( 1+\frac{2t_1}{s^{\prime}(1-\beta)}
\right) - 2 {\rm Li_2} \left( 1+ \frac{s^{\prime}}{2t_1} (1+\beta) \right)
\nonumber \\
&+& 2 {\rm Li_2} \left( 1+\frac{2u_1}{s^{\prime}(1-\beta)}
\right) - 2 {\rm Li_2} \left( 1+ \frac{s^{\prime}}{2u_1} (1+\beta) \right)
\nonumber \\
&+& \ln^2 \left[ -\frac{s^{\prime}}{2t_1} (1-\beta) \right]
+ \ln^2 \left[ -\frac{s^{\prime}}{2u_1} (1-\beta) \right],
\end{eqnarray}
and
\begin{eqnarray}
\label{SQED}
\tilde{S}_{i,QED} &=& \frac{4}{\epsilon} \left[ \frac{1}{\beta} \left(
\frac{2m^2}{s}-1 \right) \ln \frac{1-\beta}{1+\beta} - 1 \right]
+ \frac{1}{\beta} \left( \frac{2m^2}{s}-1 \right)
\nonumber \\
& \times & \left\{
\left[ 8 \ln \tilde{\beta} -2 \ln \frac{s}{m^2}
+4 \ln \frac{s^{\prime}}{m^2} + \ln \frac{1-\beta}{1+\beta} \right]
\ln \frac{1-\beta}{1+\beta} +
4 {\rm Li_2} \left( \frac{2 \beta}{1+\beta} \right) \right\}
\nonumber \\
&-& 8 \ln \tilde{\beta} + 2 \ln \frac{s}{m^2} - 4 \ln
\frac{s^{\prime}}{m^2} - \frac{2}{\beta} \ln \frac{1-\beta}{1+\beta},
\end{eqnarray}
where $\zeta(2) = \pi^2 / 6$.

%
%
\newpage
\mysection*{Appendix C}
\setcounter{section}{3}

In this appendix we discuss the derivation of the collinear limit $y=-1$
of the matrix element $M_i^g(3)$.  A similar analysis follows for
$M_i^{q}(3)$ for which we simply quote the results at the end.
We closely follow appendix B of \cite{mnr} thereby deriving
$f_i^g(x,-1,\theta_1,\theta_2)$ in $n$ dimensions
(see also appendix C of \cite{ks}).

Consider the diagrams shown in fig.\ 3.
The momenta are defined according to fig.\ 1.
Take the incoming gluon to have color index $a$ and Lorentz index $\mu$,
the outgoing gluon to have color index $c$ and Lorentz index $\sigma$,
and the photon to have Lorentz index $\rho$.  Then the
sum of the diagrams is
\begin{equation}
\label{Mfirst}
M^{ac \rho} = \epsilon^{\mu}(k_1) \bar{\epsilon}^{\sigma}(k_2)
\left[ D_{\mu \sigma}^{ac \rho} + R_{\mu \sigma}^{ac \rho} \right] \, ,
\end{equation}
where $D_{\mu \sigma}^{ac \rho}$ is the contribution from fig.\ 3(d) and
$R_{\mu \sigma}^{ac \rho}$ represents the remaining terms that
are regular when $y=-1$.  Using standard Feynman rules one finds
\begin{equation}
D_{\mu \sigma}^{ac \rho} = \left[ -if^{acb}V_{\mu \sigma \alpha}
(k_1,-k_2,k_2-k_1) \right]
\left[ \frac{-\delta^{bd} P^{\alpha}_{\beta}
(k_1-k_2)}{(k_1-k_2)^2} \right] B^{d \beta \rho}(q,k_1-k_2) \, ,
\end{equation}
with
\begin{equation}
V_{\mu \sigma \alpha}(k_1,-k_2,k_2-k_1) = (k_1+k_2)_{\alpha} g_{\mu \sigma}
+ (k_1-2k_2)_{\mu} g_{\sigma \alpha} + (k_2-2k_1)_{\sigma} g_{\mu \alpha} \, ,
\end{equation}
and $B^{b \alpha \beta}$ the sum of the lowest order diagrams in fig.\ 2.
We choose the propagator in the light-like axial gauge where
\begin{equation}
P_{\mu \nu}(k) = -g_{\mu \nu} +
\frac{ k_{\mu} \eta_{\nu} + k_{\nu} \eta_{\mu}}{\eta \cdot k} \, .
\end{equation}
Now $\eta$ and $k_1$ define a plane.  Take $k_{\perp}$ perpendicular to
this plane and choose $\eta$ such that $\eta \cdot k_1 \neq 0$ and
$\eta^2=0$.  Then decomposing $k_2=(1-x)k_1+\eta \xi + k_{\perp}$ implies
that $\xi = -k_{\perp}^2 / [ (1-x)2 \eta \cdot k_1]$ and
$t^{\prime} = k_{\perp}^2 / (1-x)$.  The collinear limit we require is
$k_{\perp} \rightarrow 0$ with $x$ fixed.
If we use $\epsilon(k_1) \cdot k_1 =0$, $\bar{\epsilon}(k_2) \cdot k_2 =0$,
and $(k_1-k_2)_{\alpha} P^{\alpha}_{\beta}(k_1-k_2)=O(k_{\perp}^2)$ then
\begin{equation}
V^{\mu \sigma \alpha} (k_1,-k_2,k_2-k_1) \simeq \frac{2}{x} k_{\perp}^{\alpha}
g^{\mu \sigma} -2 k_{\perp}^{\mu} g^{\sigma \alpha} + \frac{2}{1-x}
k_{\perp}^{\sigma} g^{\mu \alpha} + \cdots + O(k_{\perp}^2) \, ,
\end{equation}
where $\cdots$ represents terms that vanish upon contraction with the
polarization tensors.  Hence (\ref{Mfirst}) becomes
\begin{eqnarray}
M^{ac \rho} &=& \left\{ -igf^{acb} \left[ \frac{2}{x} k_{\perp \alpha}
g_{\mu \sigma} -2 k_{\perp \mu} g_{\sigma \alpha} + \frac{2}{1-x}
k_{\perp \sigma} g_{\mu \alpha} + \cdots + O( k_{\perp}^2)  \right]
\right. \nonumber \\
& \times & \left. \left[
\frac{-\delta^{bd} P_{\beta}^{\alpha}(k_1-k_2)}{(k_1-k_2)^2} \right]
B^{d \beta \rho}(q,k_1-k_2) + R_{\mu \sigma}^{ac \rho}
\right\} \epsilon^{\mu} (k_1)
\bar{\epsilon}^{\sigma} (k_2) \, . \nonumber \\
&&
\end{eqnarray}
{}From this result one can see that
$M \sim O( 1/ \sqrt{t^{\prime}} ) + R + O( t^{\prime} )$ as
$t^{\prime} \rightarrow 0$.  Hence when we square $M$
we can drop interference terms containing $R$
because only the square of the first
term gives a $1/ t^{\prime}$ singularity.  Now because
$k_1-k_2 = xk_1+O(k_{\perp})$ we have
$B^{a \beta \rho}(q,k_1-k_2)=B^{a \beta \rho}(q,xk_1)+O(k_{\perp})$
so we can write
\begin{eqnarray}
\label{Afinal}
M^{ac \rho} \left( M^{ac \lambda} \right)^{\dagger} &=&
- \frac{4g^2C_A}{t^{\prime}} B^{a \beta \rho}(q,xk_1)
\left( B^{a \alpha \lambda}(q,xk_1) \right)^{\dagger}
P_{\alpha \beta}(k_1) \nonumber \\
& \times & \left[ (1-x) + \frac{1}{1-x} + \frac{1-x}{x^2} \right]
\nonumber \\
&+& \frac{4g^2C_A}{t^{\prime}} B^{a \beta \rho}(q,xk_1)
\left( B^{a \alpha \lambda}(q,xk_1) \right)^{\dagger}
\frac{(1-x)(n-2)}{x^2} \nonumber \\
& \times & \left[ \frac{k_{\perp \alpha} k_{\perp \beta}}{k_{\perp}^2} +
\frac{P_{\alpha \beta}}{n-2} \right] \, .
\end{eqnarray}
Using the definition of $f^g_i(x,y,\theta_1,\theta_2)$ in eq.\ (\ref{fgi})
one can easily make the identification used in the main text, namely that
\begin{equation}
f^g_i(x,-1,\theta_1,\theta_2) = f^g_i(x,\theta_1)
+ \tilde{f}^g_i(x,\theta_1,\theta_2).
\end{equation}
The second term vanishes upon integration over $\theta_2$ and
\begin{eqnarray}
f^g_i(x,\theta_1) &=& 512 \pi^2 \mu^{-2 \epsilon} \alpha_s^2 e^2 e^2_H N
C_A C_F \frac{(s^{\prime})^2}{s} \nonumber \\ &\times&
\frac{1-x}{x} \left[
x(1-x) + \frac{x}{1-x} + \frac{1-x}{x} \right] B_{i,QED}(xk_1) .
\end{eqnarray}
The mass parameter $\mu$ originates from the
dimensionality of the gauge coupling in $n$ dimensions.
For the quark channel a similar analysis holds with the result
\begin{equation}
f^q_i(x,-1,\theta_1,\theta_2) = f^q_i(x,\theta_1)
+ \tilde{f}^q_i(x,\theta_1,\theta_2)
\end{equation}
where again the second term vanishes upon integration over $\theta_2$ and
\begin{eqnarray}
f^q_i(x,\theta_1) &=& -128 \pi^2 \mu^{-2 \epsilon}
\alpha_s^2 e^2 e^2_H N C_F (1+\epsilon/2)^{-1} \nonumber \\ & \times &
\left[ \frac{1+(1-x)^2+ \epsilon x^2/2}{x^2} \right] B_{i,QED} (xk_1).
\end{eqnarray}

Note that we need the full $f_i^g(x,-1,\theta_1,\theta_2)$ in $n=4$ dimensions
in the plus distributions used in the main text not just $f_i^g(x,\theta_1)$.
For this one may take the $n=4$ limit of eq.\ (\ref{Afinal}) by
using an explicit representation of $k_{\perp}$.
However, we found it instructive to take the collinear limit of
$M_i^g(3)$ in the Feynman gauge and compare with the axial gauge results.
It is well known \cite{hump} that if one chooses a gauge different
from the axial one in QCD, then interference graphs also become
singular in the collinear limit.
The axial gauge result follows from the previous analysis.
In the Feynman gauge the squared matrix element arising from the
diagrams in fig.\ 3 (including interference terms) has the form
\begin{equation}
\label{Mexp}
M_i^g(3) = \frac{A}{(t^{\prime})^2} + \frac{B}{t^{\prime}}
+ C + {\cal O}( t^{\prime} ),
\end{equation}
where $A$, $B$, and $C$ are functions of the other nine Mandelstam invariants
defined in appendix A.  Expanding the invariants $u_1$, $u_7$, and $t_1$ about
$t^{\prime}=0$ we find
\begin{eqnarray}
\label{colin}
u_1 &=& u_1^c - \frac{t^{\prime}}{s^{\prime}_5} \left[ u_1^c
+ Q^2 \beta_5 \cos \theta_1 \right] \, , \nonumber \\
u_7 &=& x t_1^c - \frac{t^{\prime}}{s^{\prime}_5} \left[ x t_1^c
- Q^2 \beta_5 \cos \theta_1 \right] \, , \nonumber \\
t_1 &=& t_1^c - \frac{s^{\prime}}{2} \sqrt{2 a t^{\prime}} \beta_5
\cos \theta_2 \sin \theta_1 \nonumber \\ &+& \frac{t^{\prime}}{s^{\prime}_5}
\left\{ u_1^c+\beta_5 \cos \theta_1 \left[ Q^2 \left(
\frac{x-1}{x} \right) + s^{\prime} \right] \right\} \, ,
\end{eqnarray}
where
\begin{equation}
a = -2 (1-x) \frac{s_5}{(xs^{\prime})^2} \, ,
\end{equation}
and
\begin{eqnarray}
t_1^c &=& - \frac{1}{2} s^{\prime}   ( 1 - \beta_5 \cos \theta_1 )
\, , \nonumber \\
u_1^c &=& - \frac{1}{2} s_5^{\prime} ( 1 + \beta_5 \cos \theta_1 ).
\end{eqnarray}
Direct substitution of the invariants (\ref{colin}) into (\ref{Mexp}) and
using the definition of $f^g_G(x,y,\theta_1,\theta_2)$ in eq.\
(\ref{fgi}) yields
the results
\begin{eqnarray}
\label{fgGcol}
f^g_G(x,-1,\theta_1,\theta_2) &=& -
\frac{32(2m^2-Q^2)(xs^{\prime})^2}{(xt_1^c)u_1^c s}
g(x,\theta_1,\theta_2) \nonumber \\
&+& \frac{32(x^2-x+1)^2(xs^{\prime})^2}{x^4s}
\left[ \frac{(xs^{\prime})^2}{(xt_1^c) u_1^c} -2 \right] \, , \nonumber \\ & &
\end{eqnarray}
\begin{equation}
\label{fgLcol}
f^g_L(x,-1,\theta_1,\theta_2) = -\frac{64Q^2}{s} g(x,\theta_1,\theta_2) \, ,
\end{equation}
where
\begin{eqnarray}
g(x,\theta_1,\theta_2) &=& \frac{2(x^2-2x+2)}{x^2} \left[ m^2
\frac{(xs^{\prime})^2}{(xt_1^c)u_1^c} -s_5 \right] \nonumber \\
&-& \frac{(1-x)^2}{x^4} s_5 \beta_5^2 \cos^2 \theta_2 \sin^2 \theta_1
\frac{(xs^{\prime})^2}{(xt_1^c)u_1^c} \, . \nonumber \\ & &
\end{eqnarray}
Both $f^g_G(x,-1,\theta_1,\theta_2)$ and $f^g_L(x,-1,\theta_1,\theta_2)$
have a common factor $g^4e^2e^2_HNC_AC_F$ which we have not shown.
We have used the relation $xs^{\prime}+xt_1^c+u_1^c=0$ freely in
deriving (\ref{fgGcol}) and (\ref{fgLcol}).
A similar analysis for the quark channel gives the results
\begin{eqnarray}
f^q_G(x,-1,\theta_1,\theta_2) &=& \frac{16(2m^2-Q^2)}{(xt_1^c)u_1^c}
h(x,\theta_1,\theta_2) \nonumber \\
&-& \frac{8(x^2-2x+2)}{x^2} \left[ \frac{(xs^{\prime})^2}{(xt_1^c) u_1^c}
-2 \right] \, , \nonumber \\ & &
\end{eqnarray}
\begin{equation}
f^q_L(x,-1,\theta_1,\theta_2) = \frac{32Q^2}{(xs^{\prime})^2}
h(x,\theta_1,\theta_2) \, ,
\end{equation}
where
\begin{equation}
h(x,\theta_1,\theta_2) = m^2
\frac{(xs^{\prime})^2}{(xt_1^c)u_1^c} -s_5 - \frac{(1-x)}{x^2} s_5 \beta_5^2
\cos^2 \theta_2 \sin^2 \theta_1 \frac{(xs^{\prime})^2}{(xt_1^c)u_1^c} \, .
\end{equation}
Both $f^q_G(x,-1,\theta_1,\theta_2)$ and $f^q_L(x,-1,\theta_1,\theta_2)$
have a common factor $g^4e^2e^2_HNC_F$ which we have not shown.
We have checked the formulae for the $f^{a_1}_i(x,-1,\theta_1,\theta_2)$
$(a_1=g,q, \, \, i=G,L)$ presented above
by comparing them numerically with the $y \rightarrow -1$ limit of the
complete expression $f^{a_1}_i(x,y,\theta_1,\theta_2)$.


\newpage
\centerline{Figure Captions}

\begin{description}

\item[fig.1.]
Illustration of the basic reaction (\protect{\ref{reaction}})
for heavy-flavor production in virtual-photon-parton
collisions.
\item[fig.2.]
Lowest order Feynman diagram contributing to the amplitude for
the gluon fusion reaction (\protect{\ref{born}}).  Another diagram
is obtained by reversing the arrows on the heavy quark lines.
\item[fig.3.]
The order $eg^2$ diagrams contributing to the amplitude for the
gluon-bremsstrahlung reaction (\protect{\ref{gbrem}}).  Additional
graphs are obtained by reversing the arrows on the heavy quark lines.
\item[fig.4.]
One loop order $eg^3$ diagrams for the virtual corrections
contributing to the reaction (\protect{\ref{born}}).  Except for the
non-planar diagram additional diagrams are obtained by reversing the
arrows on the heavy quark lines. The solid line in the gluon self-energy
graph represents gluon, ghost and quark loops.
\item[fig.5.]
The order $eg^2$ diagrams contributing to the amplitude for the
reaction (\protect{\ref{quark}}).  Additional graphs are obtained
by reversing the arrows on the light-quark lines (dashed).
\item[fig.6.]
The $\eta$ dependence of the scaling function $c^{(1)}_{L,q}(\eta,\xi)$
in the $\overline{{\rm MS}}$ scheme for several values of $Q^2$ (in units
of $({\rm GeV/c})^2$ ) with $m=4.75$ (GeV/$c^2$), {\em i.\ e.\ } the corrected
version of fig.\ 9(b) of ref.\ \protect\cite{LRSvN1}.
The solid line corresponds to $Q^2=0.01$, the dotted line to $Q^2=1$,
the short dashed line to $Q^2=10$, the long dashed line to $Q^2=100$, and the
dot-dashed line to $Q^2=1000$.
\item[fig.7.]
(a) The $\eta$ dependence of the scaling function
$d^{(1)}_{T,q}(\eta,\xi)$ in the $\overline{{\rm MS}}$ scheme
for several values of $Q^2$ (in units of $({\rm GeV/c})^2$ )
with $m=4.75$ (GeV/$c^2$), {\em i.e.} the corrected
version of fig.\ 11(a) of ref.\ \protect\cite{LRSvN1}.
The short dashed line corresponds to $Q^2=10$, the long dashed line to
$Q^2=100$, and the dot-dashed line to $Q^2=1000$.
(b) The $\eta$ dependence
of the scaling function $d^{(1)}_{L,q}(\eta,\xi)$ in the
$\overline{{\rm MS}}$ scheme for the same values of $Q^2$ with
$m=4.75$ (GeV/$c^2$) {\em i.e.} the corrected version of fig.\ 11(b) of
ref. \protect\cite{LRSvN1}.
\item[fig.8.]
(a) The distributions $dF_2(x,Q^2,m_c^2,P_t)/dP_t$ for
charm-anticharm pair production at fixed $Q^2=12 \, ({\rm GeV/c})^2$
with $x=$ $4.2 \times 10^{-4}$ (solid line), $8.5 \times 10^{-4}$
(dotted line), $1.6 \times 10^{-3}$ (short dashed line) and
$2.7 \times 10^{-3}$ (long dashed line). (b) The same distributions at
fixed $x=8.5 \times 10^{-4}$ and $Q^2=$ $8.5$ (solid line),
$12$ (dotted line), $25$ (short dashed line),
$50$ (long dashed line) all in units of $({\rm GeV/c})^2$.
\item[fig.9.]
(a) The distributions
$dF_2(x,Q^2,m_c^2,\Delta \phi)/d(\Delta \phi)$ for charm-anticharm
pair production at the $x$ and $Q^2$ values given in fig.\ 8(a).
(b) The same distributions at the $x$ and $Q^2$ values given in fig.\ 8(b).
\item[fig.10.]
(a) The distributions $dF_2(x,Q^2,m_c^2,R)/dR$ for charm-anticharm
pair production at the $x$ and $Q^2$ values given in fig.\ 8(a).
(b) The same distributions at the $x$ and $Q^2$ values given in fig.\ 8(b).
\item[fig.11.]
(a) The distributions $dF_L(x,Q^2,m_c^2,P_t)/dP_t$ for
charm-anticharm pair production at the $x$ and $Q^2$ values given
in fig.\ 8(a).  (b) The same distributions at the $x$ and $Q^2$
values given in fig.\ 8(b).
\item[fig.12.]
(a) The distributions $dF_L(x,Q^2,m_c^2,\Delta \phi)/d(\Delta \phi)$
for charm-anticharm pair production at the $x$ and $Q^2$ values given
in fig.\ 8(a).  (b) The same distributions at the $x$ and $Q^2$
values given in fig.\ 8(b).
\item[fig.13.]
(a) The distributions $dF_L(x,Q^2,m_c^2,R)/dR$ for charm-anticharm
pair production at the $x$ and $Q^2$ values given
in fig.\ 8(a).  (b) The same distributions at the $x$ and $Q^2$
values given in fig.\ 8(b).
\item[fig.14.]
(a)  The distributions $dF_2(x,Q^2,m_b^2,P_t)/dP_t$ for
bottom-antibottom pair production at the $x$ and $Q^2$ values given
in fig.\ 8(a).  (b) The same distributions at the $x$ and $Q^2$
values given in fig.\ 8(b).
\item[fig.15.]
(a)  The distributions $dF_2(x,Q^2,m_b^2,\Delta \phi)/d(\Delta \phi)$
for bottom-antibottom pair production at the $x$ and $Q^2$ values given
in fig.\ 8(a).  (b) The same distributions at the $x$ and $Q^2$
values given in fig.\ 8(b).
\item[fig.16.]
(a)  The distributions $dF_2(x,Q^2,m_b^2,R)/dR$ for
bottom-antibottom pair production at the $x$ and $Q^2$ values given
in fig.\ 8(a).  (b) The same distributions at the $x$ and $Q^2$
values given in fig.\ 8(b).
\item[fig.17.]
(a)  The distributions $dF_L(x,Q^2,m_b^2,P_t)/dP_t$ for
bottom-antibottom pair production at the $x$ and $Q^2$ values given
in fig.\ 8(a).  (b) The same distributions at the $x$ and $Q^2$
values given in fig.\ 8(b).
\item[fig.18.]
(a)  The distributions
$dF_L(x,Q^2,m_b^2,\Delta \phi)/d(\Delta \phi)$ for bottom-antibottom
pair production at the $x$ and $Q^2$ values given
in fig.\ 8(a).  (b) The same distributions at the $x$ and $Q^2$
values given in fig.\ 8(b).
\item[fig.19.]
(a)  The distributions $dF_L(x,Q^2,m_b^2,R)/dR$ for
bottom-antibottom pair production at the $x$ and $Q^2$ values given
in fig.\ 8(a).  (b) The same distributions at the $x$ and $Q^2$
values given in fig.\ 8(b).
\end{description}

%
%

\end{document}